\documentclass[pra, reprint, superscriptaddress, amsmath, amssymb]{revtex4-1}

\usepackage{graphicx}

\usepackage{dcolumn}
\usepackage{bm}

\usepackage{amssymb, amsmath}
\usepackage{textcomp}
\usepackage{wasysym}
\usepackage{mathrsfs}

\newtheorem{theorem}{Theorem}
\newtheorem{lemma}{Lemma}
\newtheorem{definition}{Definition}

\begin{document}
	
	\title{Randomized Benchmarking Using Unitary $t$-Design for Average Fidelity Estimation of Practical Quantum Circuit}
	
	\noaffiliation
	\author{Linxi Zhang}
	\affiliation{State Key Laboratory of Integrated Services Networks, Xidian University, Xi'an, Shannxi, China}

	\author{Chuanghua Zhu}
	\email{chhzhu@xidian.edu.cn}
	\affiliation{State Key Laboratory of Integrated Services Networks, Xidian University, Xi'an, Shannxi, China}
	
	\author{Changxing Pei}
	\affiliation{State Key Laboratory of Integrated Services Networks, Xidian University, Xi'an, Shannxi, China}
	
	\date{\today}

	\begin{abstract}
		Randomized benchmarking is a useful scheme for evaluation the average fidelity of a noisy quantum circuit. However, it is insensitive to the unitary error. Here, we propose a method of randomized benchmarking in which a unitary $t$-design is applied and by which the unitary error estimation can be converted to analysis of pseudo-randomness on a set of unitary operators. We give a bound on the number of randomized benchmarking sequences, when performing a unitary $t$-design on $n$-qubit $d$-dimensional system. By applying local random unitary operators, a decomposition of a unitary $t$-design, the bound is more practical than the previous bound for multi-qubit circuit. We also give a rigorous bound of a diamond norm between arbitrary and uniform distributions of a set of unitary operators to form an $\epsilon$-approximate unitary $t$-design. It can be used to quantitatively analyze the corresponding average fidelity and errors in a large-scale quantum circuit. 
	\end{abstract}
	
	\maketitle

	\newcommand{\0}{\text{\textrangle}}
	\newcommand{\9}{\text{\textlangle}}
	\newcommand{\textbardb}{\textnormal{\textbardbl}}
	

	\section{Introduction}
	
	To implement fault-tolerant quantum computation \cite{Got2009_Gottesman} is to design highly reliable quantum circuits. However, there always exists errors for various noise. To remove or decrease the error, we must characterize the errors of quantum circuits first. Use of quantum process tomography (QPT) \cite{DPS2003AiIaEPV1p22_DAriano, GG2003_Gill, PCZ1996_Poyatos} for error characterization is feasible. In QPT, a known quantum state is used to probe an unknown quantum process to obtain a description of the circuit. These methods can be divided into two types: direct and indirect \cite{MRL2007PRA702_Mohseni}.
	
	In indirect method \cite{NC2010_Nielsena, Leu2002JMPV4N22p5_Leung, DP2000PRLv8p42_DAriano, ABJ+2003Prl_Altepeter, DP2002PRL902_DAriano}, the QPT is derived from quantum state tomography \cite{AJK2004_Altepeter}. Generally, the known input state and the corresponding output results are used to estimate the transfer matrix. However, the indirect QPT approach to noise estimation suffers from several practical deficiencies. Firstly, it is sensitive to state preparation and measurement (SPAM). Secondly, the number of experiments grows exponentially in accordance with increasing the number of qubits and dimensions. Finally, in each experiment, we must calculate an approximated result iteratively; the accuracy of this result is limited by the high computational complexity.
	
	On the other hand, in the direct method, experimental results rather than the transfer matrix directly provide the required information. There has been growing interest in the development of direct QPT for diagnosing noise in quantum circuits, through methods such as direct fidelity estimation \cite{dSLP2011PRL122_Silva, FL2011PRL122_Flammia}, gate set tomography \cite{BGN+2016NC82_Blume-Kohout, BGN+2013_Blume-Kohout, MGS+2012PRA802_Merkel}, and randomized benchmarking \cite{EAZ2005JOBQSO72S_Emerson, KLR+2007_Knill, MGE2010PRL112_Magesana, MGE2011PRA802_Magesan, GMT+2012PRL122_Gaebler, MGJ+2012PRL102_Magesan, WF2014NJP121_Wallman, WBE2014NJP102_Wallman, WE2015PRA902_Wallman, HWFW2017_Helsen}. Direct fidelity estimation is a simple method for providing an estimate of the fidelity between a desired quantum process $\Lambda$ and the actual result obtained in the laboratory, up to a constant additive error. However, direct fidelity estimation, which gives an unconditional estimate of the average gate fidelity, is susceptible to SPAM. Gate set tomography provides a full and accurate tomographic description of every gate with confidence bounds. However, although it eliminates the impact of SPAM, considerable resources are required. Randomized benchmarking measures an average error rate that closely estimates the average process infidelity. It is also insensitive to SPAM; thus, it is an efficient method for estimation of the coherence of channel noise \cite{WGHF2015NJP112_Wallman, WBE2015PRL102_Wallman}. However, it is insensitive to unitary errors \cite{SWS2015NJoP102_Sanders, BGN+2016NC82_Blume-Kohout} and, therefore, cannot measure diamond norm error \cite{KLDF2015PRL112_Kueng} precisely. 
	
	In randomized benchmarking, the average survival probability \cite{PRY+2017PRL112_Proctor} over all randomized benchmarking sequences $C_{s_i}$ of length $r$ is 
	\begin{equation}
		P_r=\frac{1}{|\mathcal{C}|^r}\sum_s tr(EC_{s^{-1}}C_{s_r}\cdots C_{s_1}\rho),
	\end{equation}
	where $E$ is the measurement operator, $\rho$ is the input state and $C_{s^{-1}}$ is the Clifford gate, which satisfies $C_{s^{-1}}C_{s_r}\cdots C_{s_1}=|\mathcal{C}|^r\mathbb{I}$. The motivation is that the unitary invariance of the Fubini-Study measure make us to turn any Fubini-Study integral into an integral over the Haar measure on $\mathcal{U}(d^n)$. The average fidelity $\mathbb{E}(F(\psi,U))$ with an input state $|\psi\0$ and set of unitary operators can be defined as
	\begin{equation}
	\begin{split}
	& \mathbb{E}_{|\psi\0}(F(\psi,U)) \equiv \mathbb{E}_{\{U\}}(F(\psi_0,U))\\
	=&\int_{F-S}F(|\psi\0,U)d\psi\\
	=&\int_{\mathcal{U}(d^n)}F(|\psi_0\0,U)\mu_{Haar}(dU), 
	\end{split}
	\end{equation}
	where the integration of F-S is with respect to the Fubini-Study measure, $\mu_{Haar}(dU)$ denotes the Haar measure, and $|\psi_0\0$ denotes an arbitrary input state. Because the uniform distribution over the Clifford group is an exact unitary $2$-design \cite{DCEL2006PRA802_Dankert}, the average survival probability $P_r$ of a sequence of Clifford gates can be efficiently estimated \cite{WF2014NJP121_Wallman}. 
	
	However, how to construct a multi-qubit Clifford group is currently unknown. For the number of qubit $n=1$, we generally use Clifford gates (i.e. CNOT gates, Hadamard gates and $\pi/4$ phase gates) to construct a Clifford group in randomized benchmarking. For $n\geq 2$, it is not efficient to construct a Clifford group on $n$ qubits for unitary 2-design. Hence, we consider the randomized benchmarking sequences $C_{s_i}$ as a polynomial $t$-design (defined in Definition \ref{definitionp}). It is equal to the definition of a unitary $t$-design. The advantage of this method is to efficiently construct a random circuit for multi-qubit system.
	
	In this paper, we extend the unitary $t$-design concept,a scheme by which the unitary group is distributed to imitate the characterization of the Haar measure for polynomials of degree up to $t$ \cite{BHH2012_Brandao}, for error estimation of a large-scale quantum circuit. It can also be represented in the tensor power $U^{\otimes t}\otimes (U^\dagger)^{\otimes t}$ of the representation $U$ and its conjugate. There are also many applications of the unitary $t$-design. For example, $t=1$ corresponds to the case of a private quantum channel \cite{AMTdW2000_Ambainis}, whereas $t=4$ corresponds to the conditions of the state-distinction problem \cite{Sen2005_Sen, AE2007_Ambainis}. Following the extension of the unitary $t$-design, we use local random unitary operator to construct an $\epsilon$-approximate unitary $t$-design, which can be used to analyze the diamond norm error between the arbitrary and uniform distributions of the unitary group. By applying this method, we convert unitary error analysis in practice into an investigation of the pseudo-randomness of the unitary group distribution, which can performed through the analysis of the random walk.

	We show below that an exact unitary $t$-design can be constructed using local random unitary operators over a Clifford group on a Markovian noise channel for average fidelity estimation. It is an efficient way to construct the randomized benchmarking sequences for multi-qubit system. We extend the proof of Dankert $et$ $al.$  \cite{DCEL2006PRA802_Dankert}, i.e., that the uniform distribution over the Clifford group is a unitary $2$-design, to a general $t$ by applying a specific decomposition of the local random unitary operators. We then extend the bound of BHH \cite{BHH2012_Brandao} and prove the connection of the spectral gap with a random walk of $n$ steps. We give a bound on the number of sequences required, when performing a unitary $t$-design on $n$-qubit $d$-dimensional system, to obtain an $\epsilon$-approximate unitary $t$-design for arbitrary distribution of a set of unitary operators. Finally, we apply an $\epsilon$-approximate unitary $t$-design to a kind of linear large-scale quantum circuit to estimate the average fidelity and errors caused by practical implementation.
	
	\section{Definitions and results}
	
	We derive a definition of a polynomial $t$-design as follows, based on Refs. \cite{ELL2005PRA702_Emerson, DCEL2006PRA802_Dankert}.
	
	\begin{definition}\label{definitionp}
		Let $\nu$ be a distribution over a finite set $\{U_m\}_{m=1}^M\subset \mathcal{U}(d^n)$ on $n$-qubit $d$-dimensional systems. Define a polynomial $t$-design for every polynomial $P_{(t,t)}(U)$ of degree at most $t$ in the matrix elements of $U$, and at most $t$ in the complex conjugates of those matrix elements,
		\begin{equation}
			\frac{1}{M}\sum_{m=1}^{M}P_{(t,t)}(U_m)=\int_{\mathcal{U}(d^n)}P_{(t,t)}(U)\nu(dU).
		\end{equation}
	\end{definition}
	
	The above definition gives a method in which the average polynomial over a finite set of unitary operators is rated to an integral with a kind of distribution over $\mathcal{U}(d^n)$. In randomized benchmarking, a sequence of monomials over a set of unitary operators is generally used to estimate the average infidelity. Meanwhile, a unitary $t$-design is a distribution of a set of unitary operators which mimic properties of the Haar measure for polynomial $t$-design in the tensor power $U^{\otimes t,t}$ of the fundamental representation $U$ and its conjugate. Therefore, a unitary $t$-design can be expressed in terms of quantum operations as follows.
	
	\begin{definition}
		Let $\Lambda$ be a general quantum channel on $n$-qubit $d$-dimensional quantum state $\rho$. There exists a $\nu$-twirl process that transforms the superoperator $\Lambda$ to the superoperator $\Delta_{\nu}^t$ through application of unitary $t$-design. The resulting superoperator satisfies
		\begin{equation}
			\begin{split}
			\Delta_{\nu}^t: \rho \mapsto & \int_{\mathcal{U}(d^n)}U^{\otimes t,t}\Lambda (U^\dagger)^{\otimes t,t}\rho\nu(dU)\\
			= & \int_{\mathcal{U}(d^n)}U^{\otimes t}\Lambda((U^\dagger)^{\otimes t}\rho U^{\otimes t})(U^\dagger)^{\otimes t}\nu(dU)
			\end{split}
		\end{equation}
	\end{definition}

	From the above definitions, we derive a strong definition of the $\epsilon$-approximate unitary $t$-design from Ref. \cite{BHH2012_Brandao}. If $\mathcal{N}_1$, $\mathcal{N}_2$ are superoperators on $n$-qubit $d$-dimensional systems and $|\Phi_{d^n}\0=d^{-n/2}\sum_{i=1}^{d^n}|i,i\0$ is the maximally entangled state, we can define a symbol $\preceq$ as the usual semidefinite ordering such that $\mathcal{N}_1\preceq \mathcal{N}_2$ satisfies
	\begin{equation}
		(\mathcal{N}_1\otimes \mathbb{I})|\Phi_{d^n}\0 \preceq (\mathcal{N}_2\otimes \mathbb{I})|\Phi_{d^n}\0.
	\end{equation}

	\begin{definition}
		Let $\nu$ and $\mu_{Haar}$ be the arbitrary and uniform Haar distributions on $\mathcal{U}(d^n)$, respectively. Then, the $\epsilon$-approximate unitary $t$-design satisfies
		\begin{equation}
			\begin{split}
			(1-\epsilon)(\Delta_{\mu_{Haar}}^t\otimes\mathbb{I})|\Phi_{d^n}\textnormal{\0} & \preceq (\Delta_{\nu}^t\otimes \mathbb{I})|\Phi_{d^n}\textnormal{\0} \\
			& \preceq (1+\epsilon)(\Delta_{\mu_{Haar}}^t\otimes\mathbb{I})|\Phi_{d^n}\textnormal{\0}. 
			\end{split}
		\end{equation}
	\end{definition}

	To simplify the definition, we define an $\epsilon$-approximate unitary $t$-design in terms of the diamond norm i.e.,
	\begin{equation}
	\textbardb \Delta_{\nu}^t-\Delta_{\mu_{Haar}}^t \textbardb_\Diamond \leq 2\epsilon,
	\end{equation}
	where the diamond norm of a superoperator $\Delta$ is defined as in Ref. \cite{BH2010QIC102_Brandao}:
	\begin{equation}
	\textbardb \Delta \textbardb_\Diamond=\sup_d \textbardb \Delta\otimes \mathbb{I}_d \textbardb_{1\rightarrow 1}.
	\end{equation}
	Here, the $p\rightarrow q$ induced Schatten norm is $\textbardb \Delta(X) \textbardb_{1\rightarrow 1}=\sup_{X\neq 0}\frac{\textbardb \Delta(X) \textbardb_p}{\textbardb X \textbardb_q}$. The diamond norm is generally used as the quantity to prove the fault-tolerance thresholds \cite{AB2008SJoC_Aharonov}. Moreover, the diamond norm can be used to indicate the worst-case error rate, where
	\begin{equation}
		\epsilon(\Lambda)=\frac{1}{2}\textbardb \Lambda-\mathbb{I} \textbardb_\Diamond=\sup_{|\Phi_{d^n}\0}\textbardb (\Lambda\otimes\mathbb{I}_{d^n}-\mathbb{I}_{d^{2n}}) \textbardb_1.
	\end{equation}
	
	There is no specific experimental process corresponding to the diamond norm. Therefore, we must scale the bound of the results. However, the complexity of a fidelity estimation increases with the number of qubits $n$. Therefore, it is difficult to find a general solution to directly scale the bound. Here, we follow the approach of Brandao and Horodecki in Ref. \cite{BH2010QIC102_Brandao}, using a local random circuit to construct a unitary $t$-design combined with the properties of the random walk to scale the bound.
	
	\begin{definition}\label{DefinitionLocal}
		(Local random unitary operators) In each step of the walk, an index $i$ is chosen uniformly at random from the set $\{1,\cdots,n\}$. A two-qubit unitary $U_{i,i+1}$ drawn from a set of Haar measures $\mathcal{U}(d^n)$ is applied to the two neighboring qubits $i$ and $i+1$. (Because of the finite numbers of the qubits, we arrange the $(n+1)$-th qubit as being equal to the first qubit.)
		
		The operator $H_{n,t}$ is a quantum local Hamiltonian composed of local normalized operators $H_{i,i+1}$ of ordered neighboring subsystems, such that
		\begin{equation}
		H_{n,t}=\frac{1}{n}\sum_{i=1}^n H_{i,i+1},
		\end{equation} 
		with local terms $H_{i,i+1}=\mathbb{I}-P_{i,i+1}$, Here, $P_{i,i+1}$ is the projector of two neighbors $i$, $i+1$, on $\mathcal{U}(d^n)$, such that
		\begin{equation}
		P_{i,i+1}=\int_{\mathcal{U}(d^n)}{U_{i,i+1}}^{\otimes t,t}\mu_{Haar}(dU).
		\end{equation}
	\end{definition}

	The properties of local random unitary operators are as follows \cite{BH2010QIC102_Brandao, BV2009_Brown}:
	\begin{itemize}
		\item (Periodic boundary conditions) The (n+1)-th subsystems is identified with the first.
		\item (Zero ground-state energy) $\lambda_{min}(H_{n,t})=0$, with $\lambda_{min}(H_{n,t})$ being the minimum eigenvalue of $H_{n,t}$.
		\item (Frustration-freeness) Every state $|\psi\0$ in the groundstate manifold, composed of all eigenvectors with eigenvalue zero, is such that $H_{i,i+1}|\psi\0=0$, for all $i\in \{1,\cdots,n\}$.
	\end{itemize}
	
	Our first contribution is to extend the results of Refs. \cite{DLT2001ITITV4N352_DiVincenzo, DCEL2006PRA802_Dankert}.
	
	\begin{theorem}\label{Theoremtdesign}
		A unitary $t$-design can be constructed using a local random unitary operator with a uniform distribution over the Clifford group acting on a completely positive linear superoperator.
	\end{theorem}

	It is difficult to provide a representation of a large complex multi-qubit system. Therefore, we use the model of local random unitary operators to consider the interaction between two ordering qubits only. Through this approach, the complexity of fidelity estimation is reduced to a great extent. Although this method limits the construction of the  practical implementation, it is efficient for obtaining a result to estimate the average fidelity of a quantum process.
	
	Our second contribution is to extend the proof of the results in Ref. \cite{BHH2012_Brandao} rigorously, considering $n$ steps of a random walk.
	
	\begin{theorem}\label{TheoremRW}
		Consider a unitary $t$-design formed by local random unitary operators with size $r$. Let $\nu$ and $\mu_{Haar}$ be the arbitrary and uniform distributions on $\mathcal{U}(d^n)$, respectively. The diamond norm of the $\epsilon$-approximate unitary $t$-design over a Markov channel $\Lambda$ on $(n+1)$-qubit $d$-dimensional systems satisfies
		\begin{equation}\label{Equation2t}
			\textbardb \Delta_{\nu}^t-\Delta_{\mu_{Haar}}^t \textbardb_\Diamond \leq (2t)^{\frac{1}{nr}}C^{\frac{1}{n^2}},
		\end{equation}
		where
		\begin{equation}
			C=1-\frac{1}{e^{n}(d^2+1)^{n-1}}.
		\end{equation}
	\end{theorem}

	Because of the normalized local random unitary operator, the diamond norm satisfies
	\begin{equation}\label{Equation1}
		\textbardb \Delta_{\nu}^t-\Delta_{\mu_{Haar}}^t \textbardb_\Diamond \leq 1.
	\end{equation}
	Inserting inequality (\ref{Equation2t}) into inequality \ref{Equation1}, we obtain
	\begin{equation}\label{Equationr}
		r\geq \frac{n\ln \frac{1}{2t}}{\ln C}.
	\end{equation}
	
	\begin{figure}[!htb]
		\centering
		\includegraphics[scale=0.5]{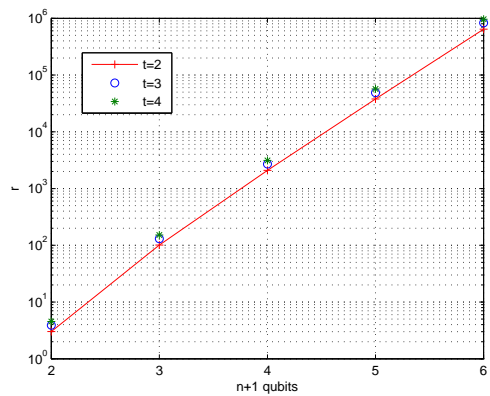}
		\caption{\label{figurer1} Size $r$ vs number of qubits $n$. For illustration purpose, we assume $d=2$}
	\end{figure}
	
	From FIG. \ref{figurer1}, we note that the size of a unitary $t$-design for an arbitrary distribution exhibits exponential growth with increasing $n$. If we take a unitary $t$-design with size larger than $r$ in randomized benchmarking, we can estimate the error caused by pseudo-randomness. In other words, an $\epsilon$-approximate unitary $t$-design with an arbitrary distribution can be used to estimate the unitary error, when its size is larger than $r$. Compared with the factor $n$, the change of $t$ has little effect on $r$.

	We can further improve the bound of the diamond norm by relating it to the spectral gap of the local random unitary operator. The aim is to reduce the influence of $n$. Although the error is increased, this approach greatly improves the estimation efficiency.
	
	Our third contribution is to extend the bound of Ref. \cite{BHH2012_Brandao}.

	\begin{theorem}\label{TheoremBound}
		An $\epsilon$-approximate unitary $t$-design in terms of the diamond norm on $n$-qubit $d$-dimensional systems satisfies
		\begin{equation}
			\begin{split}
			&\textbardb \Delta_{\nu}^t-\Delta_{\mu_{Haar}}^t \textbardb_\Diamond \\
			\leq &1-(e^2(d^2+1)[2t(t-1)]^{3}\lceil0.8\log_d[2t(t-1)]+1 \rceil^2+1)^{-1}.
			\end{split}
		\end{equation}
		
	\end{theorem}

	\begin{figure}[!htb]
		\centering
		\includegraphics[scale=0.5]{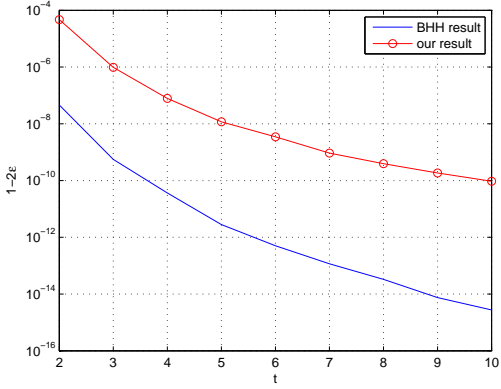}
		\caption{\label{figured2} Relationship between $1-\textbardb \Delta_{\nu}^t-\Delta_{\mu_{Haar}}^t \textbardb_\Diamond$ and $t$. For illustration, we assume $d=2$. The blue line represents the result of the $\epsilon$-approximate unitary $t$-design given in Ref. \cite{BHH2012_Brandao}. To obtain an approximate result, we let $r$ approach infinity and take $n$ as having minimum value of 2. The red line corresponds to simulated results obtained for $r\geq \frac{n\ln \frac{1}{2t}}{\ln C}$.}
	\end{figure}

	From FIG. \ref{figured2}, arbitrary distribution $\nu$ yields an error approaching 1. However, we can use the scheme in Ref. \cite{TM2015PRL122_Turner} to construct the unitary $t$-design in an experiment. Then, we can estimate the error obtained for a real distribution by applying this method. Hence, we convert the unitary error to an investigation of the pseudo-randomness of a set of unitary operator distributions. The error caused by the experimental pseudo-randomness has a considerable influence on the estimation of the result.
	
	In theory, we establish a boundary for the range of the diamond norm of a given $\epsilon$-approximate unitary $t$-design. In practice, we can recognize the diamond as a quality factor by applying a unitary $t$-design. We can evaluate the randomness by varying the number of unitary selections, which is dependent on only three factors $n$, $r$ and $t$. From FIG. \ref{figured2}, it is apparent that use of small $t$ and large $r$ for estimation of the average fidelity is preferable. However, increased $r$ creates greater complexity with regard to constructing a quantum circuit. Decreased $t$ makes greater needs of unique unitary operators to construct a random circuit in practical randomized benchmarking. Further, from FIG. \ref{figurer1}, the change of $t$ has little effect on $r$. Therefore, it is preferable to use the smallest $r$ and largest $t$ to fit the experimental requirements.

\begin{figure}[!htb]
		\centering
		\includegraphics[scale=0.35]{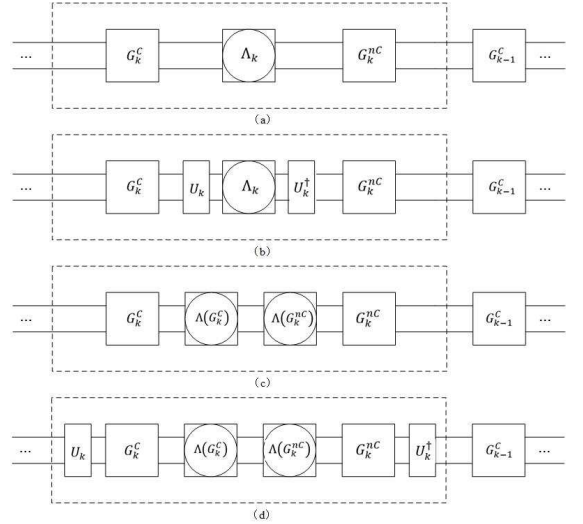}
		\caption{\label{figures2} (a) Fragment of noisy gate-independent circuit, with $k$-th round indicated by dashed box. $G_k^C$ represents a non-trivial Clifford gate and $G_k^{nC}$ is the general designation of a non-Clifford and gate $\mathbb{I}$. $\Lambda_k$ represents $k$-th round of the completely positive trace-preserving (CPTP) gate-independent noise channel. (b) Unitary $t$-design acting on $k$-th CPTP noise channel. (c) Fragment of noisy gate-dependent circuit with the $k$-th round represented by the dashed box. $\Lambda(G_k^C)$ and $\Lambda(G_k^{nC})$ represent the noise channels dependent on gates $G_k^C$ and $G_k^{nC}$, respectively. (d) Unitary $t$-design acting on $k$-th round of circuit.}
	\end{figure}

	
	Following the noise circuit in Ref. \cite{WE2015PRA902_Wallman}, we construct a quantum circuit that consists of $K$ rounds of subsystems. using a model of a single-qubit unitary 1-design to tailor the gates, Wallman and Emerson separated the circuit into easy and hard gates. However, in the fragment considered here, which is presented in FIG. \ref{figures2}, we use local random unitary operators to construct a unitary $t$-design that can be applied in a kind of large-scale quantum circuit. We separate the circuit into gates constructed using non-trivial Clifford gates and non-Clifford gates. (Note that we add gate $\mathbb{I}$ to the group of non-Clifford gates and use non-Clifford gates as a general designation.) We also denote the noise channel in the circuit as a trace-preserving channel to neglect certain attenuation parameters.
	
	We first consider a Markovian noise channel independent of the gates in the circuit shown in FIG. \ref{figures2} (a). The circuit consists of only $K$-rounds of subsystems, including a Clifford gate $G_k^C$, a non-Clifford gate $G_k^{nC}$, and a gate-independent channel $\Lambda_k$. The fidelity for the input state $|\psi\0$ and a group of quantum algorithms $\{U_j\}$ follows
	\begin{equation}
		\begin{split}
		& F(|\psi\0,\{U_j\})\\
		=& tr(\rho_{|\psi\0} G_K^C U_K \Lambda_K U_K^\dagger G_K^{nC}\cdots G_1 U_1 \Lambda_1 U_1^\dagger G_1^{nC} \rho_{|\psi\0} ).
		\end{split}
	\end{equation}
	
	Our fourth contribution is a fidelity estimation application, performed using an $\epsilon$-approximate unitary $t$-design.
	\begin{theorem}\label{TheoremFidelity}
		We consider a quantum circuit $C$ consisting of only $K$-rounds of subsystems arranged linearly, including a Clifford gate $G_k^C$, a non-Clifford gate $G_k^{nC}$ (including the trivial gate $\mathbb{I}$), and a completely positive trace-preserving (CPTP) gate-independent Markovian noise channel $\Lambda_k$. If we take a unitary $t$-design on each of the subsystems, the $\varepsilon(U^{\otimes t,t})$-approximate average fidelity of the circuit $C$ satisfies
		\begin{equation}
		\overline{F}(C)\leq \textbardb \prod_{i=1}^K G_i^C G_i^{nC}\textbardb_1 \frac{tr(\prod_{j=1}^K \Delta_{\mu_{Haar},j}^t)+d^n}{d^{2n}+d^n}+2\varepsilon(U^{\otimes t,t}),
		\end{equation}
		where
		\begin{equation}
		\varepsilon(U^{\otimes t,t})= F_g(\rho_{|0\0},\prod_{i=1}^K G_i^C G_i^{nC})\cdot \sum_{j=1}^K \epsilon_j.
		\end{equation} 
		$F_g(\rho_{|0\0}, \prod_{i=1}^K G_i^C G_i^{nC})$ denotes the gate fidelity between $\rho_{|0\0}$ and the gate $\prod_{i=1}^K G_i^C G_i^{nC}$.
	\end{theorem}
	
	Because it is difficult to find a way to prove that a non-Clifford gate can constitute a unitary $t$-design, we cannot directly estimate the fidelity of the gate-dependent noise channel circuit in FIG. \ref{figures2} (c). Instead of analyzing the specific noise channel model, however, we can calculate the difference between the gate-dependent and gate-independent noise channel to estimate the approximate range of the results.
	
	Our fifth contribution is the following theorem:
	\begin{theorem}\label{TheoremDifference}
		Let $C_{GD}$ and $C_{GI}$ be circuits to which unitary $t$-designs are applied, having gate-dependent and gate-independent trace-preserving noise channels, respectively. The difference of the diamond norm between these two circuits satisfies
		\begin{equation}
			\begin{split}
			& \textbardb C_{GD}-C_{GI} \textbardb_\Diamond \\
			\leq & \textbardb \prod_{l=1}^K G_l^C \textbardb_1\textbardb \prod_{m=1}^K G_m^{nC} \textbardb_1 \sum_{k=1}^K(\textbardb 	\Lambda(G_k^C)\Lambda(G_k^{nC})-\Lambda_k \textbardb_\Diamond\\
			&+2\varepsilon_{dif,k}),
			\end{split}
		\end{equation}
		where
		\begin{equation}
			2\varepsilon_{dif,k}= 1-\frac{1}{\textbardb G_k^{nC}\textbardb} F_g(\rho_{|\psi\0},G_k^{nC}) + 2\epsilon_k.
		\end{equation}
	\end{theorem}
	
	\section{Methods}
	
	\subsection{Proof of Theorem \ref{Theoremtdesign}}
	
	We consider a completely positive superoperator $\Lambda$ linear mapping expressed as $\Lambda(\rho)=A\rho B$ only, where $A,B\in L(\mathbb{C}(d^n))$. Our approach is to construct a unitary $t$-design by local random unitary operators on a subset of the Clifford group $\mathcal{C}(d^n)$. Note that the generalized Pauli group $\mathcal{P}(d^n)$, which consists of all $n$-fold tensor products of the one-qubit Pauli operators $\{\mathbb{I},X,Y,Z\}$, is a normal subgroup of $\mathcal{C}(d^n)$. Therefore, it is sufficient to consider the sympathetic group $\mathcal{S}L(d^n)=\mathcal{C}(d^n)/\mathcal{P}(d^n)$ \cite{Cha2004_Chau}. 
	
	WE define
	\begin{equation}
	\begin{split}
	&Sum_{d^{tn}}(\rho,\Lambda)\\
	=&\frac{1}{|\mathcal{C}(d^{tn})|}\sum_{U\in \mathcal{C}(d^n)}M_{(t,t)}(U)\Lambda(M_{(t,t)}^\dagger(U)\rho M_{(t,t)}(U))M_{(t,t)}^\dagger(U),
	\end{split}
	\end{equation}
	and
	\begin{equation}
	\begin{split}
	& Int_{d^{tn}}(\rho,\Lambda)\\
	=& \int_{\mathcal{U}(d^{tn})} U^{\otimes t}\Lambda((U^\dagger)^{\otimes t}\rho U^{\otimes t})(U^\dagger)^{\otimes t}\mu_{Haar}(dU),
	\end{split}
	\end{equation}
	where $M_{(t,t)}(U)$ is the random polynomial circuit consisting of the unitary operators.(Generally, we often make $M_{(t,t)}(U)$ as a random monomial circuit normalized as a unitary operator on $\mathcal{C}(d^{tn})$.) Our approach is to prove that the above two equalities are equal.
	
	We can use the local random unitary operators to simplify the above definition. In practice, we can construct a parallel random unitary operator on $n$ qubits using the even and odd unitary operators \cite{TM2015PRL122_Turner}. In this paper, however, we only explicitly consider the original proposal in Definition \ref{DefinitionLocal}. We redefine
	\begin{equation}
	\begin{split}
	& Sum_{d^{tn}}(\rho,\Lambda)\\
	=& \frac{1}{|\mathcal{C}(d^{tn})|}\sum_{M_{n,t}\in \mathcal{C}(d^{tn})}M_{n,t}\Lambda(M_{n,t}^\dagger \rho M_{n,t})M_{n,t}^\dagger,
	\end{split}
	\end{equation}
	and
	\begin{equation}
	\begin{split}
	& Int_{d^{tn}}(\rho,\Lambda)\\
	= & \frac{1}{n}\sum_{i=1}^n\int_{\mathcal{U}(d^n)} U_{i,i+1}^{\otimes t}\Lambda((U_{i,i+1}^\dagger)^{\otimes t}\rho U_{i,i+1}^{\otimes t})(U_{i,i+1}^\dagger)^{\otimes t}\\
	&\cdot \mu_{Haar}(dU_{i,i+1}),
	\end{split}
	\end{equation}
	where $M_{n,t}=1/n\sum_{i=1}^n M_{i,i+1}$ and 
	\begin{equation}
	M_{i,i+1}=\int_{\mathcal{U}(d^n)}U_{i,i+1}^{\otimes t}\mu_{Haar}(dU_{i,i+1}).
	\end{equation}
	
	As shown in Ref. \cite{EAZ2005JOBQSO72S_Emerson}, Schur's lemma (for a detailed derivation and application, please refer to Ref. \cite{GW2009_Goodman}) implies that $\hat{\Lambda}$ is a $\mathcal{U}(d^{tn})$-invariant trace-preserving superoperator given by
	\begin{equation}
	\begin{split}
	\hat{\Lambda}(\rho)&=p\rho+(1-p)Tr(\rho)\frac{\mathbb{I}}{d^{tn}},\\
	& = \int_{\mathcal{U}(d^n)} U^{\otimes t}\Lambda((U^\dagger)^{\otimes t}\rho U^{\otimes t})(U^\dagger)^{\otimes t}\mu_{Haar}(dU),
	\end{split}
	\end{equation}
	where
	\begin{equation}
	p=\frac{tr(\hat{\Lambda})-1}{d^{2tn}-1}.
	\end{equation}
	
	Now, we consider the completely positive channel $\Lambda(\rho)=A\rho B$ combined with Schur's lemma, such that
	\begin{equation}
	\begin{split}
	&Int_{d^n}(\rho,\Lambda)\\
	=& \frac{1}{n}\sum_{i=1}^n \int_{\mathcal{U}(d^n)} U_{i,i+1}^{\otimes t} A (U_{i,i+1}^\dagger)^{\otimes t}\rho U_{i,i+1}^{\otimes t} B (U_{i,i+1}^\dagger)^{\otimes t}\\
	&\cdot\mu_{Haar}(dU_{i,i+1}),\\
	=& \frac{d^{tn}tr(A)tr(B)-tr(AB)}{d^{tn}(d^{2tn-1})}\rho+\frac{d^{tn}tr(AB)-tr(A)tr(B)}{d^{tn}(d^{2tn}-1)}tr(\rho)\mathbb{I}.
	\end{split}
	\end{equation}
	
	Then, we denote the elements of $\mathcal{P}(d^{tn})$ as $\{P_j\}_{j=1}^{d^{2tn}}$, where $P_1$ is the $nt$-fold tensor product of $\mathbb{I}$. We can define $A=\sum_{a=1}^{d^{tn}}\alpha_aP_a$, $B=\sum_{b=1}^{d^{tn}}\beta_bP_b$, and $\rho=\sum_{j=1}^{d^{tn}}\gamma_jP_j\rho P_j$. The expression of the Pauli-twirled superoperator is given by
	\begin{equation}
	\begin{split}
	\Lambda_P(\rho)= & \frac{1}{d^{2tn}}\sum_{j=1}^{d^{2tn}}P_jAP_j^\dagger\rho P_jBP_j^\dagger,\\
	=& \frac{1}{d^{2tn}}\sum_{a=1}^{d^{2tn}}\sum_{b=1}^{d^{2tn}}\alpha_a\beta_b\sum_{j=1}^{d^{2tn}}\omega^{(j,a-b)_{Sp}}P_a\rho P_b^\dagger,\\
	=& \sum_{a=1}^{d^{2tn}}\alpha_a\beta_a P_1 \rho P_a,
	\end{split}
	\end{equation}
	where
	\begin{equation}
	\sum_{j=1}^{d^{2tn}} \omega^{(j,a-b)_{Sp}}=d^{2tn}\delta_{a,b}.
	\end{equation}
	From the above equalities, the $\mathcal{S}L(d^{tn})$-twirl yields
	\begin{equation}
	\begin{split}
	& \Lambda_U(\rho)\\
	= & \frac{1}{|\mathcal{S}L(d^{tn})|}\sum_{M_{n,t}\in \mathcal{S}L(d^{tn})}M_{n,t}\Lambda_P(M_{n,t}^\dagger \rho M_{n,t})M_{n,t}^\dagger,\\
	= & \frac{|\mathcal{P}(d^{tn})|}{|\mathcal{C}(d^{tn})|}\sum_{M_{n,t}\in \mathcal{S}L(d^{tn})}\sum_{j=1}^{d^{2tn}}\alpha_j\beta_jM_{n,t}P_jM_{n,t}^\dagger\rho M_{n,t}P_j^\dagger M_{n,t}^\dagger,\\
	=& (\alpha_1\beta_1-\frac{1}{d^{2tn}-1}(\sum_{j=2}^{d^{2tn}}\alpha_j\beta_j))\rho+\frac{d^{2tn}}{d^{2tn}-1}(\sum_{j=2}^{d^{2tn}}\alpha_j\beta_j)tr(\rho)\mathbb{I}.
	\end{split}
	\end{equation}
	The definition of $\mathcal{C}(d^{tn})=\mathcal{S}L(d^{tn})\circ \mathcal{P}(d^{tn})$ implies
	\begin{equation}
	Sum_{d^{tn}}(\rho,\Lambda)=Int_{d^{tn}}(\rho,\Lambda).
	\end{equation}
	
	
	\subsection{Proof of Theorem \ref{TheoremRW}}
	
	Following the work of Brandao, Harrow, and Horodecki \cite{BHH2012_Brandao}, we further prove the connection of the spectral gap of $H_{n,t}$ with the random walk. $H_{n,t}$ is defined as the local random unitary operators on $n$ $d$-dimensional systems. We use the convergence time of the random walk \cite{Gur2016_Guruswami, BD1997_Bubley} to lower the bound of the spectral gap.
	
	We consider an intermediate notion of convergence to the random walk based on the $L^p$ Wasserstein distance between two probability measures $\mu$ and $\nu$ \cite{Oli2007AoAP2V1N31_Oliveira}. The $L^p$ Wasserstein distance is expressed as
	\begin{equation}
	W_p(\nu,\mu):=\inf\{(\mathbb{E}[d(U_\nu,U_\mu)^p])^{\frac{1}{p}}\},
	\end{equation}
	where $(U_\nu,U_\mu)$ is a pair of random variables coupling $(\nu,\mu)$. We note that
	\begin{equation}
	W(\nu,\mu)=W_1(\nu,\mu)\leq W_2(\nu,\mu).
	\end{equation}
	
	In this paper, we use $\int fd\nu$ to estimate the fidelity of the system for any Lipschitz $f$, such that
	\begin{equation}
	\begin{split}
	W_1(\nu,\mu):=\sup\{ & \int_{\mathcal{U}(d)}f(U)\nu(dU)-\int_{\mathcal{U}(d)}f(U)\mu(dU):\\
	& f: \mathcal{U}(d)\rightarrow \mathbb{R} \text{ is } 1-\text{ Lipschitz under } d\}.
	\end{split}
	\end{equation}
	
	\begin{lemma}\label{Lemmadistance}
		(Ref. \cite{Oli2007AoAP2V1N31_Oliveira}) A map $f:M\rightarrow N$ between Polish length spaces ($M,d_M$) and ($N,d_N$) is said to be locally $C$-Lipschitz (for some $C>0$) if, for all $x,y\in M$, 
		\begin{equation}
		\limsup_{y\rightarrow x} \frac{d_N(f(x),f(y))}{d(x,y)}\leq C.
		\end{equation}
		Suppose $P$ is a Markov transition kernel on ($M,d$) satisfying $x\rightarrow P_x$ from ($M,d$) to ($Pr_p(M),W_p$), is locally $C$-Lipschitz. Moreover, $P$ is locally $C$-Lipschitz on $M$.
		
		Then, for all $\nu_1,\nu_2\in Pr_p(M)$, we also have $\nu_1P,\nu_2P\in Pr_p(M)$,and
		\begin{equation}
		\frac{W_2(\nu_1P,\nu_2P)}{W_2(\nu_1,\nu_2)}\leq C.
		\end{equation}
	\end{lemma}	
	
	\begin{lemma}\label{LemmaRW}
		For every integer $n>1$, there exists an $n$-fold local random Hamiltonian distribution $(\nu_{LR,n+1,d})^{\ast n}$, which implies that $n$ steps of a random walk acting on $(n+1)$-qubit systems are taken, satisfying
		\begin{equation}
		W((\nu_{LR,n+1,d})^{\ast n},\mu_{Haar})\leq(1-\frac{1}{e^{n}(d^2+1)^{n-1}})^{\frac{1}{n}}\sqrt{2}d^{\frac{n+1}{2}}.
		\end{equation}
	\end{lemma}
	\textbf{Proof:} We use mathematical induction to prove the above inequality.
	
	(1) Firstly, for $n=2$, it is shown in Ref. \cite{BHH2012_Brandao} that two steps of a random walk on three-qubit systems establish the inequality. (The specific proof is given in Appendix \ref{AppendixTwosteps}.)
	
	(2) Then, consider $n-1$ steps of a random walk on $n$-qubit systems satisfying Lemma \ref{LemmaRW}. Let $R_{1,n-1}$ and $R_{2,n-1}$ be two unitary operators on $n$-qubit $d$-dimensional systems. For $n-1$ steps of the walk, there are $(n-1)^{n-1}$ possibilities, each occurring with probability $1/(n-1)^{n-1}$; therefore,
	\begin{equation}
		R_{2,n-1}\rightarrow R_{2,n-1}'=U_{i_{n-1},i_{n-1}+1}\cdots U_{i_1,i_1+1}R_{2,n-1}.
	\end{equation}
	We also have $n!$ possibilities of a non-trivial coupling for which 
	\begin{equation}
		R_{1,n-1}\rightarrow R_{1,n-1}'=U_{i_{n-1},i_{n-1}+1}V_{i_{n-1},i_{n-1}+1}\cdots U_{i_1,i_1+1}R_{1,n-1},\\
	\end{equation}
	where the unitary operator $V_{i_{n-1},i_{n-1}+1}$ depends on $U_{i_j,i_j+1}$, with $j\text{ and }i_j\in\{1,\cdots,n-2\}$. The average of two infinitesimally close unitaries is given as 
	\begin{equation}
	\begin{split}
	&\inf_{V_{i_{n-1},i_{n-1}+1}}\mathbb{E}(\textbardb R_{1,n-1}'- R_{2,n-1}' \textbardb_2^2)\\
	=& 2(tr(\mathbb{I})-\mathbb{E}(\textbardb tr_{i_1\cdots i_{n-2}} (U_{i_{n-2},i_{n-2}+1}\cdots U_{i_1,i_1+1}\\
	& \cdot R_{1,n-1}R_{2,n-1}^\dagger U_{i_{n-2},i_{n-2}+1}^\dagger\cdots U_{i_1,i_1+1}^\dagger)\textbardb_1)),\\
	=& \varepsilon^2[tr(H^2)-\frac{1}{d^{n-2}}\mathbb{E}(tr(( tr_{i_1\cdots i_{n-2}} (U_{i_{n-2},i_{n-2}+1}\cdots U_{i_1,i_1+1}\\
	&\cdot R_{1,n-1}R_{2,n-1}^\dagger U_{i_{n-2},i_{n-2}+1}^\dagger\cdots U_{i_1,i_1+1}^\dagger))^2))]+O(\varepsilon^3),\\
	\leq & \varepsilon^2(1-\frac{1}{(d^2+1)^{n-2}})tr(H^2)+O(\varepsilon^3).
	\end{split}
	\end{equation}
	
	(3) Finally, we consider $n$ steps of a random walk acting on $(n+1)$-qubit systems. Let $R_{1,n}$ and $R_{2,n}$ be two unitary operators on $(n+1)$-qubit $d$-dimensional systems. We consider the same transformation as defined above, i.e.,
	\begin{equation}
	\begin{split}
	& R_{1,n}\rightarrow R_{1,n}',\\
	= & U_{i_n,i_n+1}V_{i_n,i_n+1}U_{i_{n-1},i_{n-1}+1}\cdots U_{i_1,i_1+1}R_{1,n},\\
	= & U_{i_n,i_n+1}V_{i_n,i_n+1}'R_{1,n-1}',\\
	& R_{2,n}\rightarrow R_{2,n}',\\
	= & U_{i_n,i_n+1}U_{i_{n-1},i_{n-1}+1}\cdots U_{i_1,i_1+1}R_{2,n}',\\
	= & U_{i_n,i_n+1}R_{2,n-1},
	\end{split}
	\end{equation}
	where the unitary operator $V_{i_n,i_n+1}$ depends on $U_{i_j,i_j+1}$, with $j \text{ and }i_j\in\{1,\cdots,n-1\}$. Then, we define
	\begin{equation}
	V_{i_n,i_n+1}'=V_{i_n,i_n+1}U_{i_{n-1},i_{n-1}+1}V_{i_{n-1},i_{n-1}+1}^\dagger U_{i_{n-1},i_{n-1}+1}^\dagger,
	\end{equation}
	where $V_{i_n,i_n+1}'$ also depends on $U_{i_j,i_j+1}$, and $j\text{ and }i_j\in\{1,\cdots,n-1\}$.
	
	As a result of the permutation symmetry of $j\text{ and }i_j\in\{1,\cdots,n\}$, we must only compute
	\begin{equation}
	\begin{split}
	& \inf_{V_{i_n,i_n+1}}\mathbb{E}(\textbardb R_{1,n}'-R_{2,n}' \textbardb_2^2) \\
	=& \inf_{V_{i_n,i_n+1}'}\mathbb{E}(\textbardb U_{i_n,i_n+1}V_{i_n,i_n+1}'R_{1,n-1}'-U_{i_n,i_n+1}R_{2,n-1}' \textbardb_2^2),\\
	=& \varepsilon^2[ tr(H^2) - \frac{1}{d} \mathbb{E}(tr((tr_{i_n+1}(U_{i_n,i_n+1}R_{1,k}'R_{2,k}'^\dagger U_{i_n,i_n+1}^\dagger))^2))],\\
	&+O(\varepsilon^3)\\
	\leq & \varepsilon^2 (1-\frac{1}{(d^2+1)^{n-1}})tr(H^2)+O(\varepsilon^3).
	\end{split}
	\end{equation}
	From the above inequality, we are aware that $n$ steps of a random walk on $(n+1)$-qubit systems also satisfies Lemma \ref{LemmaRW}. Therefore, we can use Lemma \ref{Lemmadistance} for an $(n+1)$-qubit system to prove the inequality in Lemma \ref{LemmaRW}. \begin{flushright}
		\Square	
	\end{flushright}
	
	We have $n^{n}-n!$ paths of the walk for which we have no shrinking (as our coupling was trivial for those paths), and $n!$ paths in which we have a shrinking factor of $1-(d^2+1)^{-(n-1)}$. This gives
	\begin{equation}
	\mathbb{E}(\textbardb R_{1,n+1}'-R_{2,n+1}' \textbardb_2^2)\leq\varepsilon^2 C\textbardb R_1-R_2 \textbardb_2^2+O(\epsilon^3),
	\end{equation}
	where
	\begin{equation}
	\begin{split}
	C= & 1-\frac{n!}{n^n}\frac{1}{(d^2+1)^{n-1}},\\
	\leq & 1-\frac{1}{e^{n}(d^2+1)^{n-1}},
	\end{split}
	\end{equation}
	where we apply the bound $n!\geq n^ne^{-n}$.
	
	Applying Lemma \ref{Lemmadistance}, we have
	\begin{equation}
	\begin{split}
	&\limsup_{\varepsilon\rightarrow 0}\sup_{U_1,U_2\in \mathbb{U}(d^n)} \{ \frac{W_2((\nu_{LR,n+1,d})^{\ast n}\nu_{U_1},(\nu_{LR,n+1,d})^{\ast n}\nu_{U_2})}{\textbardb U_1-U_2 \textbardb_2}\\
	&: \textbardb U_1-U_2 \textbardb_2\leq \varepsilon\}\leq C=1-\frac{n!}{n^{n}}\frac{1}{(d^2+1)^{n-1}}.
	\end{split}
	\end{equation}
	Then, we have
	\begin{equation}
	\begin{split}
	& W((\nu_{LR,n+1,d})^{\ast nr},\mu_{Haar})\\
	= & W_2((\nu_{LR,n+1,d})^{\ast nr}\ast\delta_{\mathbb{I}},(\nu_{LR,n+1,d})^{\ast nr}\ast \mu_{Haar}),\\
	\leq & C^{\frac{r}{n}}W_2(\delta_{\mathbb{I}},\mu_{Haar}),\\
	\leq & (1-\frac{1}{e^{n}(d^2+1)^{n-1}})^{\frac{r}{n}}\sqrt{2}d^{\frac{n+1}{2}}.
	\end{split}
	\end{equation}

	\begin{lemma}\label{Lemma2t}
		(Ref. \cite{BHH2012_Brandao}) For every integer $t,d\geq 1$ and every measure $\nu$ on $\mathcal{U}(d^n)$, the normalized difference satisfies
		\begin{equation}
			\textbardb \Delta_{\nu_{LR,n,d},t}-\Delta_{\mu_{Haar},t} \textbardb_\Diamond \leq \frac{\sqrt{2}t}{d^{\frac{n}{2}}} W(\nu_{LR,n,d},\mu_{Haar}).
		\end{equation}
	\end{lemma}

	By applying Lemmas \ref{LemmaRW} and \ref{Lemma2t}, we can prove Theorem \ref{TheoremRW} directly.
	
	\subsection{Proof of Theorem \ref{TheoremBound}}
	
 	With the properties of the quantum local Hamiltonian given in Definition \ref{DefinitionLocal}, an $\epsilon$-approximate unitary 2-design in terms of the diamond norm relating to the spectral gap is given by
 	\begin{equation}
 		\textbardb \Delta_{\nu}^t-\Delta_{\mu_{Haar}}^t \textbardb_\Diamond=1-\frac{\Delta(H_{n,t})}{n}.
 	\end{equation}
 	Then, we can substantially improve the scale of the spectral gap. We consider a chain of subsystems with the local finite dimensional Hilbert space $\mathcal{H}\subset\mathbb{C}^{d^{n-m+1}}$. We also consider a family of Hamiltonians
	\begin{equation}
	H_{[m,n]}=\frac{1}{n-m+1}\sum_{i=m}^{n}H_{i,i+1}
	\end{equation}
	on $\mathcal{H}^{\otimes(n-m+1)}\subset\mathbb{C}^{d^{2(n-m+1)}}$, where the $H_{i,i+1}$ are the nearest-neighbor interaction terms, which are assumed to be projectors. $H_{[m,n]}$ consists of all interaction terms for which both subsystems belong to the interval $\{m,m+1,\cdots,n\}$. From the properties of the local Hamiltonian, the minimum eigenvalue of $H_{[m,n]}$ is 0 for all $m,n$. We define the projector of its groundstate $G_{[m,n]}$ acting on the ground space $\mathcal{G}_{[m,n]}$, i.e.,
	\begin{equation}
	\mathcal{G}_{[m,n]}=\{|\psi\0\in\mathcal{H}^{\otimes(n-m+1)}: H_{[m,n]}|\psi\0=0\}.
	\end{equation}
	
	\begin{lemma}\label{LemmaNac}
		(Ref. \cite{Nac1994CMP115_Nachtergaele}) Suppose there exist positive integers $l$ and $n_l$, and a real number $\epsilon_l\leq 1/\sqrt{l}$, such that for all $n_l\leq m \leq n$,
		\begin{equation}
		\textbardb \mathbb{I}_{A_1}\otimes G_{A_2 B}(G_{A_1A_2}\otimes \mathbb{I}_B-G_{A_1A_2B}) \textbardb_\infty\leq\epsilon_l,
		\end{equation}
		with $A_1=[1,m-l-1], A_2=[m-l,m-1],B=m$. Then,
		\begin{equation}
		\Delta(H_{1,n})\geq\Delta(H_{[1,l]})\frac{(1-\epsilon_l\sqrt{l})^2}{l-1}.
		\end{equation}
	\end{lemma}	
	
	\begin{lemma}\label{LemmaBound}
		For every integer $t$ and $n\geq2\lceil 0.8(\log_d(4\tau)+2) \rceil$,
		\begin{equation}
		\Delta(H_{n,t})\geq \frac{\Delta(H_{\lceil 0.8(\log_d(4\tau)+2)\rceil,t})}{4\lceil 0.8(\log_d(4\tau)+1) \rceil},
		\end{equation}
		where $\tau=t(t-1)/2$.
	\end{lemma}
	\textbf{Proof:} We apply Lemma \ref{LemmaNac} with $n_l=2l$ and $\epsilon_l=1/(2\sqrt{l})$. Then, for $2l\leq m\leq n$, let
	\begin{equation}
	M=\textbardb \mathbb{I}_{A_1}\otimes G_{A_2B}(G_{A_1A_2}\otimes \mathbb{I}_B-G_{A_1A_2B}) \textbardb_\infty\leq\frac{1}{2\sqrt{l}},
	\end{equation}
	with $A_1=[1,m-l-1]$, $A_2=[m-l,m-1]$, and $B=m$. Further, let
	\begin{equation}
	X_k:=\sum_{\pi\in\mathcal{S}_t}(|\psi_{\pi,d}\0\9\psi_{\pi,d}|)^{\otimes k}.
	\end{equation}
	
	To obtain the bound of $\textbardb G_{[1,k]}-X_k \textbardb_\infty$, we first calculate
	\begin{equation}
		\sum_{\pi\in\mathcal{S}_t}|\9\psi_{\sigma,d}|\psi_{\pi,d}\0|^n=\frac{(d^n+t-1)\cdots(d^n+1)d^n}{d^{tn}}.
	\end{equation}
	The specific derivation process of the above equality \cite{AE2007_Ambainis} is presented in Appendix \ref{AppendixHaarmeasure}.
	
	Now, we scale the above equality as follows:
	\begin{equation}
	\begin{split}
	\sum_{\pi\in\mathcal{S}_t}|\9\psi_{\sigma,d}|\psi_{\pi,d}\0|^n = & \prod_{i=1}^{t-1}(1+\frac{i}{d^n}),\\
	\leq & exp(\sum_{i=1}^{t-1}\frac{i}{d^n})= e^{\frac{\tau}{d^n}},
	\end{split}
	\end{equation}
	with $\tau=t(t-1)/2$. The inequality follows $\ln(1+x)\leq x$.
	
	Let $B:=\sum_{\pi\in\mathcal{S}_t}|\pi\0\9\psi_{\pi,d}|^{\otimes n}$, with $\{|\pi\0\}_{\pi\in\mathcal{S}_t}$ being an orthonormal set of vectors; then,
	\begin{equation}
	\textbardb BB^\dagger-\sum_{\pi\in\mathcal{S}_t}|\pi\0\9\pi| \textbardb_\infty\leq\sum_{\pi\neq\sigma}|\9\psi_{\sigma,d}|\psi_{\pi,d}\0|^n\leq e^{\frac{\tau}{d^n}}-1.
	\end{equation}
	Note that $BB^\dagger$ has the same eigenvalues as $B^\dagger B$. We let
	\begin{equation}
	X_{n}=\sum_{\pi\in\mathcal{S}_t}(|\psi_{\pi,d}\0\9\psi_{\pi,d}|)^{\otimes n}=B^\dagger B.
	\end{equation}
	$G_{n,t}$ is the projector onto the support of $X_{n}$, where
	\begin{equation}
	\textbardb X_{n}-G_{n,t} \textbardb_\infty\leq e^{\frac{\tau}{d^n}}-1.
	\end{equation}
	Further, we can determine the range of $X_n$, such that
	\begin{equation}
		(2-e^{\frac{\tau}{d^n}})G_{n,t}\leq X_{n}\leq e^{\frac{\tau}{d^n}}.
	\end{equation}
	
	From the above inequality, we have
	\begin{equation}
	\textbardb G_{[1,k]}-X_k \textbardb_\infty\leq e^{\frac{\tau}{d^k}}-1,
	\end{equation}
	where $\tau=t(t-1)/2$. Therefore, $\textbardb \mathbb{I}_{A_1}\otimes G_{A_2B}(G_{A_1A_2}\otimes \mathbb{I}_B-G_{A_1A_2B}) \textbardb_\infty$ is given by
	\begin{equation}\label{EquationM}
	\begin{split}
	M\leq & \textbardb \mathbb{I}_{A_1}\otimes X_{l+1}[X_{m-1}\otimes \mathbb{I}_B-X_m] \textbardb_\infty \\
	& + (e^{\frac{\tau}{d^{m-1}}}-e^{\frac{\tau}{d^m}})\textbardb \mathbb{I}_{A_1}\otimes X_{l+1} \textbardb_\infty\\
	& + (e^{\frac{\tau}{d^{l+1}}}-1)\textbardb X_{m-1}\otimes \mathbb{I}_B-X_m \textbardb_\infty \\
	& + (e^{\frac{\tau}{d^{l+1}}}-1)(e^{\frac{\tau}{d^{m-1}}}-e^{\frac{\tau}{d^m}}),\\
	\leq & \textbardb \mathbb{I}_{A_1}\otimes X_{l+1}[X_{m-1}\otimes \mathbb{I}_B-X_m] \textbardb_\infty \\
	& + e^{\frac{\tau}{d^{l+1}}}(2e^{\frac{\tau}{d^{m-1}}}-e^{\frac{\tau}{d^{m}}}-1)-(e^{\frac{\tau}{d^{m-1}}}-1).
	\end{split}
	\end{equation}
	Let
	\begin{equation}
	\begin{split}
	Y_\pi:= & \sum_{\pi\neq\sigma}(|\psi_{\sigma,d}\0\9\psi_{\sigma,d}|)^{\otimes l}(|\psi_{\pi,d}\0\9\psi_{\pi,d}|)^{\otimes l} \\
	& \otimes (|\psi_{\sigma,d}\0\9\psi_{\sigma,d}|(\mathbb{I}_B-|\psi_{\pi,d}\0\9\psi_{\pi,d}|)).
	\end{split}
	\end{equation}
	Then, the first item of inequality (\ref{EquationM}) is given by
	\begin{equation}
	\begin{split}
	& \textbardb \mathbb{I}_{A_1}\otimes X_{l+1}[X_{m-1}\otimes \mathbb{I}_B-X_m] \textbardb_\infty\\
	= & \textbardb \sum_{\pi\in\mathcal{S}_t}(|\psi_{\pi,d}\0\9\psi_{\pi,d}|)^{\otimes (m-l-1)}\otimes Y_\pi \textbardb_\infty,\\
	= & \textbardb \sum_{\pi\in\mathcal{S}_t}(B_{m-l-1}|\pi\0\9\pi|B_{m-l-1}^\dagger)\otimes Y_\pi \textbardb_\infty,\\
	\leq & \textbardb B_{m-l-1}B_{m-l-1}^\dagger \textbardb_\infty \max_\pi\textbardb Y_\pi \textbardb_\infty.
	\end{split}
	\end{equation}
	Inequality (\ref{EquationM}) can be further scaled, with
	\begin{equation}
	\begin{split}
	M \leq & e^{\frac{\tau}{d^{m-l-1}}}\max_\pi\textbardb Y_\pi \textbardb_\infty \\
	&+ e^{\frac{\tau}{d^{l+1}}}(2e^{\frac{\tau}{d^{m-1}}}-e^{\frac{\tau}{d^{m}}}-1)-(e^{\frac{\tau}{d^{m-1}}}-1),\\
	\leq & (2e^{\frac{\tau}{d^{l-1}}}-1)(e^{\frac{\tau}{d^{2l-1}}}-1).
	\end{split}
	\end{equation}
	For every $\tau<d^{l-1}$, we have
	\begin{equation}
	\begin{split}
	M \leq & (1+\frac{2\tau}{d^{l-1}}+O(\frac{2\tau^2}{d^{2l-2}}))(\frac{\tau}{d^{2l-1}}+O(\frac{\tau^2}{d^{4l-2}})),\\
	\leq & (1+\frac{2\tau}{d^l})\frac{\tau}{d^{2l-2}},\\
	\leq & \frac{2\tau}{d^{2l-2}}\leq\frac{1}{2\sqrt{l}}.
	\end{split}
	\end{equation}
	Then, choosing $l\geq\lceil 0.8(\log_d(4\tau)+2) \rceil$, we find $M\leq1/(2\sqrt{l})$.\begin{flushright}
		\Square
	\end{flushright}
	
	Lemmas \ref{LemmaRW} and \ref{Lemma2t} indicate that, for every integer $n,t$
	\begin{equation}\label{Equation71}
		\begin{split}
		& \textbardb \Delta_{\nu}^t-\Delta_{\mu_{Haar}}^t \textbardb_\Diamond=1-\frac{\Delta(H_{n,t})}{n},\\
		\leq & (2t)^{\frac{1}{r(n-1)}}(1-\frac{1}{e^{n-1}(d^2+1)^{n-2}})^{\frac{1}{(n-1)^2}}.
		\end{split}
	\end{equation}
	Inequality (\ref{Equationr}) is substituted into inequality \ref{Equation71}, we have
	\begin{equation}
		\begin{split}
		\Delta(H_{n,t})\geq & \frac{n(2t)^{\frac{1}{(n-1)^2e^{n-1}(d^2+1)^{n-1}\ln \frac{1}{2t}}}}{e[e(d^2+1)]^{n-2}(n-1)^2},\\
		\geq & \frac{n}{(n-1)^2 e^{n-1}(d^2+1)^{n-2}+1},
		\end{split}
	\end{equation}
	where the second inequality follows the Taylor expansion $e^{1+\delta}\approx e(1+\delta)$, when $\delta\rightarrow 0$.
	
	Applying Lemma \ref{LemmaBound} with $n=\lceil 0.8\log_d(2t(t-1))+2 \rceil$, we can eliminate the effects of $n$, with
	\begin{equation}
		\begin{split}
		&\Delta(H_{n,t})\\
		\geq & \frac{n}{e^2(d^2+1)[2t(t-1)]^{3}\lceil 0.8\log_d[2t(t-1)]+1 \rceil^2+1},
		\end{split}
	\end{equation}
	where we use $0.8\log_d[e(d^2+1)]\leq 3$.
	
	Finally, we can scale the range of the $\epsilon$-approximate unitary $t$-design by eliminating the influence of $n$, such that
	\begin{equation}
		\begin{split}
		&\textbardb \Delta_{\nu}^t-\Delta_{\mu_{Haar}}^t \textbardb_\Diamond \\
		\leq &1-(e^2(d^2+1)[2t(t-1)]^{3}\lceil0.8\log_d[2t(t-1)]+1 \rceil^2+1)^{-1}.
		\end{split}
	\end{equation}
	
	\subsection{Proof of Theorem \ref{TheoremFidelity} }
	
	Firstly, we consider the $j$-th round of circuit $C$. We can obtain the Haar-averaged noise superoperator with unitary $t$-design, such that
	\begin{equation}
	\Delta_{\mu_{Haar},j}^t=\int_{\mathcal{U}(d)^{\otimes n}}U_j^{\otimes t}\Lambda_j(U_j^\dagger)^{\otimes t}\mu_{Haar}(dU_j).
	\end{equation}
	Secondly, we take a distribution $\nu$ instead of uniform distribution $\mu_{Haar}$. We can obtain the $j$-th $\epsilon$-approximate unitary $t$-design over averaged noise channel, where
	\begin{equation}
	\begin{split}
	& \textbardb\Delta_{\nu,j}^t-\Delta_{\mu_{Haar},j}^t \textbardb_\Diamond\\
	= & \textbardb\int_{\mathcal{U}(d)^{\otimes n}} U_j^{\otimes t}\Lambda_j(U_j^\dagger)^{\otimes t}\nu(dU_j) \\
	& -\int_{\mathcal{U}(d)^{\otimes n}} U_j^{\otimes t}\Lambda_j(U_j^\dagger)^{\otimes t}\mu_{Haar}(dU_j)\textbardb_\Diamond,\\
	\leq & 2\epsilon_j.
	\end{split}
	\end{equation}
	Finally, we let the input state be $|0\0$. Because the gate is independent of the noise, we obtain the approximate average fidelity of the circuit $C$ as
	\begin{equation}
	\begin{split}
	& \overline{F}(C) \\
	=& tr(\rho_{|0\0}\prod_{j=1}^K [G_j^C \int_{\mathcal{U}(d)^{\otimes n}} U_j^{\otimes t}\Lambda_j (U_j^\dagger)^{\otimes t}\nu(dU_j)G_j^{nC}]\rho_{|0\0}),\\
	\leq & tr(\rho_{|0\0}\prod_{j=1}^K [G_j^C (\int_{\mathcal{U}(d)^{\otimes n}} U_j^{\otimes t}\Lambda_j (U_j^\dagger)^{\otimes t}\mu_{Haar}(dU_j)+2\epsilon_j)\\
	& \cdot G_j^{nC}]\rho_{|0\0}),\\
	\leq & tr(\rho_{|0\0} \prod_{j=1}^K [G_j^C \int_{\mathcal{U}(d)^{\otimes n}} U_j^{\otimes t}\Lambda_j (U_j^\dagger)^{\otimes t}\mu_{Haar}(dU_j)G_j^{nC}] \rho_{|0\0})\\
	& + F_g(\rho_{|0\0},\prod_{i=1}^K G_i^C G_i^{nC})\cdot 2\sum_{j=1}^K \epsilon_j,
	\end{split}
	\end{equation}
	where $F_g(\rho_{|0\0},\prod_{i=1}^K G_i^C G_i^{nC})$ represents the gate fidelity
	\begin{equation}
	\begin{split}
	F_g(\rho_{|0\0},\prod_{i=1}^K G_i^C G_i^{nC})= & (tr\sqrt{\sqrt{|0\0\9 0|}\prod_{i=1}^K G_i^C G_i^{nC} \sqrt{|0\0\9 0|}})^2,\\
	= & (tr\sqrt{|0\0\9 0|\prod_{i=1}^KG_i^C G_i^{nC}})^2.
	\end{split}
	\end{equation}
	The gate-independent noise channel can be expressed in the form
	\begin{equation}
	\begin{split}
	& tr(\rho_{|0\0} \prod_{j=1}^K [G_j^C \int_{\mathcal{U}(d)^{\otimes n}} U_j^{\otimes t}\Lambda_j (U_j^\dagger)^{\otimes t}\mu_{Haar}(dU_j)G_j^{nC}] \rho_{|0\0})\\
	= &\textbardb \prod_{i=1}^K G_i^C G_i^{nC} \textbardb_1 \frac{tr(\prod_{j=1}^K \Delta_{\mu_{Haar},j}^t)+d^n}{d^{2n}+d^n}.\\
	\end{split}
	\end{equation} 
	
	
	\subsection{Proof of Theorem \ref{TheoremDifference}}
	
	Let
	\begin{equation}
	\begin{split}
	C_{GD,k}= & U_k^{\otimes t} G_k^C \Lambda(G_k^C) \Lambda(G_k^{nC}) G_k^{nC} (U_k^\dagger)^{\otimes t},\\
	C_{GI,k}= & G_{k}^C U_k^{\otimes t} \Lambda_k (U_k^\dagger)^{\otimes t} G_k^{nC},
	\end{split}
	\end{equation}
	where $C_{GD,k}$ and $C_{GI,k}$ denote the $k$-th rounds of the $C_{GD}$ and $C_{GI}$ circuits, respectively. $G_k^C$ and $G_k^{nC}$ represent a non-trivial Clifford gate and non-Clifford gate (the non-Clifford is redefined by adding gate $\mathbb{I}$) in the $k$-th round, respectively. Then, we have
	\begin{equation}
	\begin{split}
	C_{GD}= & \int_{\mathcal{U}(d)^{\otimes n}} C_{GD,K:1}(\prod_{k=1}^K\mu_{Haar}(dU_k)),\\
	C_{GI}= & \int_{\mathcal{U}(d)^{\otimes n}} C_{GI,K:1}(\prod_{k=1}^K\mu_{Haar}(dU_k)).
	\end{split}
	\end{equation}
	The difference in the diamond trace between the gate-dependent and gate-independent circuit satisfies
	\begin{equation}\label{Cgen}
	\begin{split}
	& \textbardb C_{GD}-C_{GI} \textbardb_\Diamond\\
	\leq & \int_{\mathcal{U}(d)^{\otimes n}} \sum_{k=1}^K\textbardb C_{GD,K:k+1}(C_{GD,k}-C_{GI,k})C_{GI,k-1:1} \textbardb_\Diamond\\
	& \cdot (\prod_{k=1}^K \mu_{Haar}(dU_k)) \\
	\leq & \int_{\mathcal{U}(d)^{\otimes n}} \textbardb \prod_{l=1,l\neq k}^K G_l^C G_l^{nC}\textbardb_1 \sum_{k=1}^K\textbardb C_{GD,k}-C_{GI,k} \textbardb_\Diamond\\
	& \cdot (\prod_{k=1}^K\mu_{Haar}(dU_k)).
	\end{split}
	\end{equation}
	Now, we consider the $k$-th round of the circuit only
	\begin{equation}\label{Cgenk}
	\begin{split}
	& \textbardb C_{GD,k}-C_{GI,k} \textbardb_\Diamond\\
	= & \textbardb U_k^{\otimes t} G_k^C \Lambda(G_k^C) \Lambda(G_k^{nC}) G_k^{nC} (U_k^\dagger)^{\otimes t}\\
	& - G_{k}^C U_k^{\otimes t} \Lambda_k (U_k^\dagger)^{\otimes t} G_k^{nC}\textbardb_\Diamond,\\
	\leq & \textbardb G_k^{C} \textbardb_1 \textbardb \Lambda(G_k^C)\Lambda(G_k^{nC})-\Lambda_k \textbardb_\Diamond \textbardb G_k^{nC} \textbardb_1.
	\end{split}
	\end{equation}
	From the above inequality, by applying the Haar measure, we can obtain the diamond trace between the general circuit $C_{GD}$ and $C_{GI}$ as
	\begin{equation}
	\begin{split}
	& \textbardb C_{GD}-C_{GI} \textbardb_\Diamond \\
	\leq & \textbardb \prod_{l=1}^K G_l^C \textbardb_1\textbardb \prod_{m=1}^K G_m^{nC} \textbardb_1 \sum_{k=1}^K\textbardb \Lambda(G_k^C)\Lambda(G_k^{nC})-\Lambda_k \textbardb_\Diamond.
	\end{split}
	\end{equation}
	
	Now, we use a practical distribution $\nu$ to estimate the approximate difference between the gate-dependent and gate-independent forms of the circuit. From Eq. (\ref{Cgenk}) in the $k$-th round, we have
	\begin{equation}\label{eCgenk}
	\begin{split}
	& \textbardb C_{GD,k}-C_{GI,k} \textbardb_\Diamond\\
	\leq & \textbardb \tilde{U}_k^{\otimes t} G_k^C \Lambda(G_k^C) \Lambda(G_k^{nC}) - G_k^C U_k^{\otimes t} \Lambda_k \textbardb_\Diamond\textbardb G_k^{nC} \textbardb_1,\\
	\leq & (\textbardb \tilde{U}_k^{\otimes t} G_k^C \Lambda(G_k^C) \Lambda(G_k^{nC}) - U_k^{\otimes t} G_k^C \Lambda(G_k^C) \Lambda(G_k^{nC}) \textbardb_\Diamond\\
	& + \textbardb U_k^{\otimes t} G_k^C \Lambda(G_k^C) \Lambda(G_k^{nC}) - G_k^C U_k^{\otimes t}\Lambda_k \textbardb)\cdot\textbardb G_k^{nC} \textbardb_1,\\
	\leq & (\textbardb \frac{1}{\textbardb G_k^{nC} \textbardb^2}(G_k^{nC})^\dagger U_k^{\otimes t} G_{k}^{nC}-U_k^{\otimes t}  \textbardb_\Diamond\textbardb G_k^C \textbardb_1\\
	& +\textbardb U_k^{\otimes t}\tilde{G}_{k}^C(U_k^\dagger)^{\otimes t}\Lambda(G_k^C)\Lambda(G_k^{nC})-\tilde{G}_k^C\Lambda_k \textbardb)\cdot \textbardb G_k^{nC} \textbardb_1,\\
	\leq & (1-\frac{1}{\textbardb G_k^{nC}\textbardb} F_g(\rho_{|\psi\0},G_k^{nC}) + 2\varepsilon(\Lambda(G_k^C)\Lambda(G_k^{nC})) + 2\varepsilon(\Lambda_k)\\
	& + 2\epsilon_k )\cdot \textbardb G_k^C \textbardb_1 \textbardb G_k^{nC} \textbardb_1,
	\end{split}
	\end{equation}
	where we use $\tilde{U}_k^{\otimes t}= \frac{1}{\textbardb G_k^{nC} \textbardb^2}(G_k^{nC})^\dagger U_k^{\otimes t} G_{k}^{nC}$ in the first inequality. In the third inequality, we take $\tilde{G}_{k}^C=G_k^C U_k^{\otimes t}$. In the fourth inequality, $F_g(\rho_{|\psi\0},G_k^{nC})=(tr\sqrt{|\psi\0\9\psi|G_k^{nC}})^2$ is the gate fidelity between $\rho_{|\psi\0}$ and the $k$-th round non-Clifford gate $G_k^{nC}$, and $\epsilon_k$ denotes the error generated by the $\epsilon$-approximate unitary $t$-designs. $\varepsilon(\Lambda(G_k^C)\Lambda(G_k^{nC}))$ and $\varepsilon(\Lambda_k)$ indicate the errors of the gate-dependent and gate-independent channels, respectively.
	
	
	\section{Discussion}
	
	We can substantially improve the result for the proof of Theorem \ref{TheoremDifference} by leaving the $(k-1)$-th round inside the diamond trace in Eq. (\ref{Cgen}) and substituting Eq. (\ref{eCgenk}) for every term except $k=1$. Therefore,
	\begin{equation}\label{eCgen}
	\begin{split}
	& \textbardb C_{GD}-C_{GI} \textbardb_\Diamond\\
	\leq & \int_{\mathcal{U}(d)^{\otimes n}} \textbardb \prod_{l=1,l\neq k,k-1}G_l^C G_l^{nC} \textbardb_1 \sum_{k=1}^K \textbardb ( C_{GD,k}-C_{GI,k}) \\
	&\cdot \int_{\mathcal{U}(d)^{\otimes n}} G_m^C U_m^{\otimes t}\Lambda_m(U_m^\dagger)^{\otimes t}G_m^{nC} \nu(dU_{m=k-1}) \textbardb_\Diamond\\
	& \cdot (\prod_{k=1,k\neq m}^K\nu(dU_k))+2\varepsilon_{dif},\\
	\end{split}
	\end{equation}
	where
	\begin{equation}
	\begin{split}
	& 2\varepsilon_{dif}\\
	=& \prod_{l=1}^K(\textbardb G_l^C \textbardb_1\textbardb G_l^{nC} \textbardb_1)\sum_{k=1}^K(1-\frac{1}{\textbardb G_k^{nC}\textbardb} F_g(\rho_{|\psi\0},G_k^{nC})+2\epsilon_k).
	\end{split}
	\end{equation}
	Then, we let $\Lambda_m=(\Lambda_m-\mathbb{I})+\mathbb{I}$, such that
	\begin{equation}\label{Equation88}
	\begin{split}
	& \textbardb  ( C_{GD,k}-C_{GI,k}) \int_{\mathcal{U}(d)^{\otimes n}} G_m^C U_m^{\otimes t}\Lambda_m(U_m^\dagger)^{\otimes t}G_m^{nC}\\ &\cdot\nu(dU_{m=k-1}) \textbardb_\Diamond \\
	\leq & \textbardb ( C_{GD,k}-C_{GI,k})\int_{\mathcal{U}(d)^{\otimes n}} G_m^C U_m^{\otimes t}(\Lambda_m-\mathbb{I})(U_m^\dagger)^{\otimes t}G_m^{nC}\\
	& \cdot \nu(dU_{m=k-1})+( C_{GD,k}-C_{GI,k})\textbardb G_{k-1}^C G_{k-1}^{nC} \textbardb_1 \textbardb_\Diamond.\\
	\end{split}
	\end{equation}
	Inserting Eq. (\ref{Equation88}) this into Eq. (\ref{eCgen}), we obtain
	\begin{equation}
	\begin{split}
	& \textbardb C_{GD}-C_{GI} \textbardb_\Diamond\\
	\leq & \prod_{l=1}^K\textbardb G_l^C G_l^{nC} \textbardb_1 \textbardb C_{GD,1}-C_{GI,1} \textbardb_\Diamond\\
	& +  \int_{\mathcal{U}(d)^{\otimes n}} \prod_{l=1,l\neq k} (\textbardb G_l^C\textbardb_1 \textbardb G_l^{nC} \textbardb_1) \sum_{k=2}^K \textbardb ( C_{GD,k}-C_{GI,k}) \\
	& \cdot (\textbardb \Lambda_{k-1}-\mathbb{I} \textbardb_\Diamond+2\epsilon_{k-1}+1)\textbardb_\Diamond (\prod_{k=2}^K\nu(dU_k))+2\varepsilon_{dif},\\
	\leq & \prod_{l=1}^K (\textbardb G_l^C \textbardb_1 \textbardb G_l^{nC} \textbardb_1) \sum_{k=1}^K\textbardb \Lambda(G_k^C)\Lambda(G_k^{nC})-\Lambda_k \textbardb_\Diamond + 2\varepsilon_{dif}^{imp},\\
	\end{split}
	\end{equation}
	where
	\begin{equation}
	\begin{split}
	&2\varepsilon_{dif}^{imp}\\
	=& \prod_{l=1}^K(\textbardb G_l^C \textbardb_1\textbardb G_l^{nC} \textbardb_1) \sum_{k=1}^K (2\epsilon_{k-1}\textbardb \Lambda(G_k^C)\Lambda(G_k^{nC})-\Lambda_k \textbardb_\Diamond \\
	& + (1-\frac{1}{\textbardb G_k^{nC}\textbardb} F_g(\rho_{|\psi\0},G_k^{nC})+2\epsilon_k)(2\epsilon_{k-1}+2)).
	\end{split}
	\end{equation}
	
	
	Finally, we estimate the average fidelity of $K$ rounds of a large-scale quantum circuit.The Haar-averaged fidelity can be related to the entanglement fidelity $F_e$, which has been proposed as a means of characterizing the noise strength in a physical quantum channel $\Lambda$ \cite{Sch1996_Schumacher,DCEL2006PRA802_Dankert}. Therefore, we obtain
	\begin{equation}
	\begin{split}
	& tr(\rho_{|0\0} \prod_{j=1}^K [G_j^C \int_{\mathcal{U}(d)^{\otimes n}} U_j^{\otimes t}\Lambda_j (U_j^\dagger)^{\otimes t}\mu_{Haar}(dU_j)G_j^{nC}] \rho_{|0\0})\\
	= & \textbardb \prod_{i=1}^K G_i^C G_i^{nC} \textbardb_1 \frac{d^n F_e(\prod_{j=1}^K\Lambda_j)+1}{d^n+1},\\
	\leq & \textbardb \prod_{i=1}^K G_i^C G_i^{nC} \textbardb_1 \frac{d^n\prod_{j=1}^K F_e(\Lambda_j)+1}{d^n+1},
	\end{split}
	\end{equation}
	where $F_e(\prod_{j=1}^K\Lambda_j)$ denotes the noise strength of the entire sequence and $F_e(\Lambda_j)$ denotes the single round noise strength of the sequence, which is based on ancilla-assisted process tomography.
	
	Now, we consider the average gate fidelity error $2\varepsilon(U^{\otimes t,t})$ only, which is caused by applying the $\nu$ distribution of the unitary $t$-designs. The overall error is a linear superposition of the errors of each round. Because every unitary operator is randomly chosen from the Clifford group, this is the same as taking unitary $t$-designs over the entire sequence $C$ $K$ times. Then, we have
	\begin{equation}
	\begin{split}
	& 2\varepsilon_k(U^{\otimes t,t})\\
	= & 2\textbardb\int_{\mathcal{U}(d)^{\otimes n}} U^{\otimes t}\cdot\prod_{k=1}^K (G_k^C\Lambda_k G_k^{nC}) \cdot(U^\dagger)^{\otimes t}\nu(dU)\\
	& -\int_{\mathcal{U}(d)^{\otimes n}}  U^{\otimes t}\cdot\prod_{k=1}^K (G_k^C\Lambda_k G_k^{nC}) \cdot(U^\dagger)^{\otimes t}\mu_{Haar}(dU)\textbardb_\Diamond.\\
	\end{split}
	\end{equation}
	For all $i\in[1,K]$, we have
	\begin{equation}
	2\varepsilon(U^{\otimes t,t})=2K \max_i \varepsilon_i(U^{\otimes t,t}).
	\end{equation}
	
	\section{Conclusion}
	
	We have proven that local random unitary operators can be used to construct a unitary $t$-design for fidelity estimation. Upon application of unitary $t$-design, an error is generated by the pseudo-randomness of the distribution of the Clifford unitary operator. Therefore, we quantitatively analyzed the $\epsilon$-approximate unitary $t$-design and obtain a better bound for the arbitrary distribution of the unitary operator. From the analysis results, we can conclude that increasing $r$ is more conducive to achieving randomness and increasing $t$ is easier to construct random circuit in randomized benchmarking. The $\epsilon$-approximate unitary $t$-design is also robust against SPAM. Then, we applied this method to a large-scale quantum circuit for average fidelity estimation. Hence, the proposed approach was to be an effective tool for estimating channel noise in practice. 
	
	However, there are still some shortcomings requiring resolution. We do not known whether a non-Clifford gate can be applied to a unitary $t$-design, nor do we have knowledge of the resultant error. Further, the circuit model considered in this study was linear. Therefore, we must consider a more general circuit and find a more effective method of estimating the error. 
	
	An open question remains as to whether we can use the specific distribution of the unitary operators to estimate the error caused by unitary $t$-design application and to give a robust fault-torrent threshold of the circuit in theory. Another open question is whether we can apply our results to randomness evaluation in practice. We can use error bars \cite{FB2015PRL102_Ferrie, CR2011PRL112_Christandl, FR2015PRL102_Faist} to evaluate the experimental results and take the $\epsilon$-approximate unitary $t$-design as a quality factor acting in the confidence interval. 
	
	
	\section*{Acknowledgements}
	
	This research work was supported by the National Natural Science Foundation of China (Grant Nos. 61372076, 61701375), the 111 Project (No. B08038) and Key Research and Development Plan of Shannxi Province (No. BBD24017290001)
	
	
	%


	\appendix
	
	\section{Two steps of random walk acting on three qubit systems}\label{AppendixTwosteps}
	
	Now we will prove two steps of random walk acting on three qubit systems. Let $R_1$ and $R_2$ be two unitaries acting on three qubit $d$-dimensional systems. Consider two steps of the walk. Then we have four possibilities, each occurring with probability $1/4$,
	\begin{equation}
	R_1\rightarrow\{\tilde{U}_{12}U_{12}R_1,\tilde{U}_{23}U_{12}R_1,\tilde{U}_{12}U_{23}R_1,\tilde{U}_{23}U_{23}R_1\},
	\end{equation}
	for independent Haar distributed unitaries $U_{12},U_{23},\tilde{U}_{12},\tilde{U}_{23}$, and likewise for $R_2$. Here the different indices of the unitaries label in which different subsystems they act non-trivially.
	
	In trivial coupling, the two unitaries $R_1$ and $R_2$ must be the same transformation. Then we should consider the nontrivial coupling. We modify the transformation with the unitary $V_{23}$ and $V_{12}$, so that $\tilde{U}_{23}V_{23}$ and $\tilde{U}_{12}V_{12}$ are Haar distributed for $\tilde{U}_{23}$ and $\tilde{U}_{12}$, respectively. We have
	\begin{equation}
	\begin{split}
	R_1\rightarrow & R_1'\\
	= & \{\tilde{U}_{12}U_{12}R_1,\tilde{U}_{23}V_{23}U_{12}R_1,\tilde{U}_{12}V_{12}U_{23}R_1,\tilde{U}_{23}U_{23}R_1\}
	\end{split}
	\end{equation}
	where the unitary $V_{23}$ and $V_{12}$ depend on $U_{12}$ and $U_{23}$, respectively. And the unitary $R_2$ undergoes the same transformation as before
	\begin{equation}
	R_2\rightarrow R_2'=\{\tilde{U}_{12}U_{12}R_2,\tilde{U}_{23}U_{12}R_2,\tilde{U}_{12}U_{23}R_2,\tilde{U}_{23}U_{23}R_2\}.
	\end{equation}
	
	Then we will show the average of the distance between two unitaries $R_1$ and $R_2$ becomes closer after random walk transformation. 
	\begin{equation}
	\begin{split}
	& \mathbb{E}(\textbardb R_1'-R_2' \textbardb^2)\\
	= & \frac{1}{4}(\mathbb{E}(\textbardb \tilde{U}_{12}U_{12}R_1-\tilde{U}_{12}U_{12}R_2 \textbardb_2^2)\\
	+ & \mathbb{E}(\textbardb \tilde{U}_{23}V_{23}U_{12}R_1-\tilde{U}_{23}U_{12}R_2 \textbardb_2^2)\\
	+ & \mathbb{E}(\textbardb \tilde{U}_{12}V_{12}U_{23}R_1-\tilde{U}_{12}U_{23}R_2 \textbardb_2^2)\\
	+ & \mathbb{E}(\textbardb \tilde{U}_{23}U_{23}R_1-\tilde{U}_{23}U_{23}R_2 \textbardb_2^2)),
	\end{split}
	\end{equation}
	with the expectation taken over Haar distributed $\tilde{U_{12}},U_{12},\tilde{U_{23}},U_{23}$. We can rewrite the equation as 
	\begin{equation}
	\begin{split}
	& \mathbb{E}(\textbardb R_1'-R_2' \textbardb)= \frac{1}{4} (2\textbardb R_1-R_2 \textbardb_2^2 \\
	+ & \mathbb{E}(\textbardb V_{23}U_{12}R_1-U_{12}R_2 \textbardb_2^2)+\mathbb{E}(\textbardb V_{12}U_{23}R_1-U_{23}R_2 \textbardb_2^2)).
	\end{split}
	\end{equation}
	Since $V_{12}$ and $V_{23}$ can depend in arbitrary way on $U_{23}$ and $U_{12}$, respectively, we can take the minimum over $V_{12}$ and $V_{23}$ to get
	\begin{equation}
	\begin{split}
	& \mathbb{E}(\textbardb R_1'-R_2' \textbardb^2) \\
	= & \frac{1}{4}(2\textbardb R_1-R_2 \textbardb_2^2\\
	+ & \mathbb{E}(\min_{V_{23}}\textbardb V_{23}U_{12}R_1-U_{12}R_2 \textbardb_2^2)\\
	+ & \mathbb{E}(\min_{V_{12}}\textbardb V_{12}U_{23}R_1-U_{23}R_2 \textbardb)).
	\end{split}
	\end{equation}
	
	Then, for any two unitaries $R_1$ and $R_2$ we have
	\begin{equation}
	\textbardb R_1-R_2 \textbardb_2^2=2(tr(\mathbb{I})-Re(tr(R_1R_2^\dagger))).
	\end{equation}
	Since $R_1$ and $R_2$ are infinitesimally close we can write
	\begin{equation}
	R=R_1R_2^\dagger=e^{i\epsilon H}=\mathbb{I}+i\epsilon H-\frac{\epsilon^2}{2}H^2+O(\epsilon^3)
	\end{equation}
	for a Hermitian matrix $H$ with $\textbardb H \textbardb_2\leq 1$. Then we get
	\begin{equation}
	\begin{split}
	& \textbardb R_1-R_2 \textbardb_2^2=2(tr(\mathbb{I})-Re(tr(R_1R_2^\dagger)))\\
	= & 2(tr(\mathbb{I})-Re(tr(\mathbb{I}+i\epsilon H-\frac{\epsilon^2}{2}H^2-i\frac{\epsilon^3}{6}H^3+O(\epsilon^4))))\\
	= & \epsilon^2tr(H^2)+O(\epsilon^4)
	\end{split}
	\end{equation}
	
	Consider the term $\mathbb{E}(\min_{V_{12}}\textbardb V_{12}U_{23}R_1-U_{23}R_2 \textbardb_2^2)$ in the right side of the above equation. (The other terms have the similar results.) We have
	\begin{equation}
	\begin{split}
	& \mathbb{E}(\min_{V_{12}}\textbardb V_{12}U_{23}R_1-U_{23}R_2 \textbardb_2^2)\\
	= & 2(tr(\mathbb{I})-\mathbb{E}(\max_{V_{12}}|Re(tr(V_{12}U_{23}R_1R_2^\dagger U_{23}^\dagger))|))\\
	= & 2(tr(\mathbb{I})-\mathbb{E}\textbardb tr_3(U_{23}RU_{23}^\dagger) \textbardb_1),
	\end{split}
	\end{equation}
	where, $\textbardb X \textbardb_1=\max_{U\in\mathbb{U}}|tr(UX)|$.
	Then, we get
	\begin{equation}
	\begin{split}
	& \textbardb tr_3(U_{23}RU_{23}^\dagger)\textbardb_1\\
	= & \textbardb tr_3(\mathbb{I}+i\epsilon U_{23}HU_{23}^\dagger-\frac{\epsilon^2}{2}U_{23}H^2U_{23}^\dagger+O(\epsilon^3))\textbardb_1\\
	= & \textbardb tr_3(\mathbb{I}+i\epsilon U_{23}HU_{23}^\dagger)-\frac{\epsilon^2}{2}tr_3(U_{23}H^2U_{23}^\dagger)+O(\epsilon^3) \textbardb_1\\
	= & \textbardb tr_3(e^{i\epsilon U_{23}HU_{23}^\dagger}+\frac{\epsilon^2}{2}(U_{23}HU_{23}^\dagger)^2)-\frac{\epsilon^2}{2}tr_3(U_{23}H^2U_{23}^\dagger)+O(\epsilon^3) \textbardb_1\\
	= & tr(\mathbb{I})+\frac{\epsilon^2}{2}\frac{1}{d}tr((tr_3(U_{23}HU_{23}^\dagger))^2)-\frac{\epsilon^2}{2}tr(H^2)+O(\epsilon^3),
	\end{split}
	\end{equation}
	so that we get
	\begin{equation}
	\begin{split}
	& \mathbb{E}\inf_{V_{12}}\textbardb V_{12}U_{23}R_1-U_{23}R_2 \textbardb_2^2 \\
	= & \epsilon^2[tr(H^2)-\frac{1}{d}\mathbb{E}(tr((tr_3(U_{23}HU_{23}^\dagger))^2))]+O(\epsilon^3).
	\end{split}
	\end{equation}
	
	Now our goal is to compute the average $\mathbb{E}(tr((tr_3(U_{23}HU_{23}^\dagger))^2))$. We note that for any operator $C_{123}$ we have \cite{ADHW2006PRSA42:2_Abeyesinghe}
	\begin{equation}
	tr(C_{12}^2)=tr((C_{123}\otimes C_{\overline{123}})(\mathbb{F}_{12:\overline{12}}\otimes \mathbb{I}_{3:\overline{3}}))
	\end{equation}
	where systems with bars are copies of original systems, and $\mathbb{F}$ is the operator which swaps systems 12 with $\overline{12}$. Therefore
	\begin{equation}
	\begin{split}
	& \mathbb{E}(tr((tr_3(U_{23}HU_{23}^\dagger))^2))\\
	= & \mathbb{E}(tr((H_{123}\otimes H_{\overline{123}})(U_{23}^\dagger\otimes U_{\overline{23}}^\dagger)(\mathbb{F}_{12:\overline{12}}\otimes\mathbb{I}_{3:\overline{3}})(U_{23}\otimes U_{\overline{23}}))).
	\end{split}
	\end{equation}
	We now compute
	\begin{equation}
	\mathbb{E}((U_{23}^\dagger\otimes U_{\overline{23}}^\dagger)(\mathbb{F}_{2:\overline{2}}\otimes\mathbb{I}_{3:\overline{3}})(U_{23}\otimes U_{\overline{23}}))=\frac{d}{d^2+1}(\mathbb{I}_{23:\overline{23}}+\mathbb{F}_{23:\overline{23}}).
	\end{equation}
	Using the fact that the tensor product of swap operators is again a swap operator(e.g. $\mathbb{F}_{12:\overline{12}}=\mathbb{F}_{1:\overline{1}}\otimes\mathbb{F}_{2:\overline{2}}$), we obtain
	\begin{equation}
	\begin{split}
	& \mathbb{E}(tr((tr_3(U_{23}HU_{23}^\dagger))^2))\\
	= & \frac{d}{d^2+1}(tr((H_{123}\otimes H_{\overline{123}})\mathbb{F}_{123:\overline{123}}\\
	+ & (H_{123}\otimes H_{\overline{123}})\mathbb{F}_{1:\overline{1}}\otimes\mathbb{I}_{23:\overline{23}}))\\
	= & \frac{d}{d^2+1}(tr(H^2)+tr(H_1^2))\\
	\geq & \frac{d}{d^2+1}tr(H^2).
	\end{split}
	\end{equation}
	Hence we obtain
	\begin{equation}
	\inf_{V_{12}}\mathbb{E}(\textbardb V_{12}U_{23}R_1-U_{23}R_2 \textbardb_2^2)\leq\epsilon^2(1-\frac{1}{d^2+1})tr(H^2)+O(\epsilon^3).
	\end{equation}
	Finally, we get
	\begin{equation}
	\mathbb{E}(\textbardb R_1'-R_2' \textbardb^2)\leq\epsilon^2(1-\frac{1}{2}\frac{1}{d^2+1})tr(H^2)+O(\epsilon^3).
	\end{equation}
	
	The fact that $\int_{\mathcal{U}(d^n)}U^{\otimes t,t}\nu(dU)$ is a projector that
	\begin{equation}
	g(\nu^{\ast k},t)=g(\nu,t)^k.
	\end{equation} 
	Then, applying Lemma \ref{Lemmadistance} 
	\begin{equation}
	\begin{split}
	\limsup_{\varepsilon\rightarrow 0}\sup_{U_1,U_2\in \mathbb{U}(d^3)} \{ & \frac{W_2((\nu_{3,d})^{\ast 2}\nu_{U_1},(\nu_{3,d})^{\ast 2}\nu_{U_2})}{\textbardb U_1-U_2 \textbardb_2}:\\
	& \textbardb U_1-U_2 \textbardb_2\leq \varepsilon\}\leq C:=(1-\frac{2}{4}\frac{1}{d^2+1})^{\frac{1}{2}}.
	\end{split}
	\end{equation}
	Since $\max_{U_1,U_2}\textbardb U_1-U_2 \textbardb_2\leq\sqrt{2}d^{3/2}$ with $U_1,U_2\in\mathcal{U}(d^3)$, we get
	\begin{equation}
	\begin{split}
	W(\nu,\mu_{Haar})\leq & W_2((\nu_{3,d})^{\ast 2}\nu_{U_1},(\nu_{3,d})^{\ast 2}\nu_{U_2})\\
	\leq & (1-\frac{1}{2}\frac{1}{d^2+1})^{\frac{1}{2}}\sqrt{2}d^{\frac{3}{2}}.
	\end{split}
	\end{equation}

	\section{Haar measure of Monomials}\label{AppendixHaarmeasure}
	
	We followed the generalization of Haar average of state-component monomials in Ref. \cite{AE2007_Ambainis}.
	\begin{equation}
	\begin{split}
	\sum_{\pi\in\mathcal{S}_t}|\9\psi_{\sigma,d}|\psi_{\pi,d}\0|^n & = \frac{1}{d^{tn}}\sum_{\pi\in\mathcal{S}_t}tr(V_{d^n}(\pi)V_{d^n}(\sigma)^T)\\
	& = \frac{1}{d^{tn}}\sum_{\pi\in\mathcal{S}_t}tr(V_{d^n}(\pi\sigma^{-1}))\\
	& = \frac{1}{d^{tn}}\sum_{\pi\in\mathcal{S}_t}tr(V_{d^n}(\pi))\\
	& = \frac{t!}{d^{tn}}tr(P_{sym,t,d^n}),
	\end{split}
	\end{equation}
	with $P_{sym,t,d^n}$ the projector onto the symmetric subspace of $(\mathbb{C}^{d^n})^{\otimes t}$:
	\begin{equation}
	P_{sym,t,d^n}=\frac{1}{t!}\sum_{\pi\in\mathcal{S}_t}V_{d^n}(\pi).
	\end{equation}
	
	Consider a set of unitary operators in $d^n$-dimensional Hilbert space, $\mathcal{H}=\mathbb{C}^{d^n}$ consisting of the set of normalized pure states. These states correspond to the points of a unit sphere $\mathbb{S}^{2d^n-1}$ which is the "surface" of a ball in $2d^n$ real dimensions. There exists a unique natural measure that is induced by the invariant (Haar) measure on the unitary group $\mathcal{U}(d^n)$: a uniformly random pure state can be defined by the action of a uniformly random unitary matrix on an arbitrary reference state. The measure on pure states is distinguished by the rotational invariance of the Haar measure, which we will denote as $\mu_{Haar}(U_{|\psi\text{\textrangle}})$. Choosing a fixed representation, $U_{|\psi\text{\textrangle}}=|\psi\text{\textrangle\textlangle}\psi|=\sum_{i}c_i^2U_{|i\text{\textrangle}}$ where $|\psi\text{\textrangle}=\sum_{i}c_i|i\text{\textrangle}$. The uniform measure for normalized vectors in $\mathcal{H}$ can be expressed using the Euclidean parametrization,
	\begin{equation}
	\mu_{Haar}(dU_{|\psi\text{\textrangle}}:\psi\in\mathbb{S}^{d^n})=(\prod_{i=1}^{d^n}d^2c_i)\delta(\sum_{l=1}^{d^n}|c_l|^2-1)
	\end{equation}\\
	where $\delta$ is the Dirac delta function.
	
	The average value of any function $f:\mathcal{H}\rightarrow\mathbb{C}$ takes the explicit form,
	\begin{equation}
	\9 f(U) \0_\psi=\frac{1}{V_{\mathbb{S}^{2d^n-1}}}f(U)\mu_{Haar}(dU_{\9 \psi \0}:\psi\in\mathbb{S}^{2d^n-1}).
	\end{equation}
	Consider first calculating the volume of the unitary operators. We have
	\begin{align}
	& V_{\mathbb{S}^{2d^n-1}}=\int_{\mathbb{S}^{2d^n-1}}\mu_{Haar}(dU)\\
	= & \int_{\mathbb{S}^{2d^n-1}}(\prod_{i=1}^{d^n}d^2\frac{u_i}{r})(\frac{\sum_{l=1}^{d^n}|u_l|^2}{r^2}-1) \label{A1} \\
	= & \int_{\mathbb{S}^{2d^n-1}}(\prod_{i=1}^{d^n}d^2u_i)r^{-2d^n}(\frac{\sqrt{\sum_{l=1}^{d^n}|u_l|^2}}{r}-1) \label{A2} \\
	= & \int_{\mathbb{S}^{2d^n-1}}(\prod_{i=1}^{d^n}d^2u_i)r^{-2d^n+1}(\sqrt{\sum_{l=1}^{d^n}|u_l|^2}-r) \label{A3} ,
	\end{align}
	where we have made the change of variables $c_i=\frac{u_i}{r}$ in equality (\ref{A1}). And in the equality (\ref{A2}) we use $\delta(\sqrt{\sum_{l=1}^{d^n}|u_l|^2}-r)$ instead of $\delta(\sum_{l=1}^{d^n}|u_l|^2-r^2)$ for the variable radius $r^2=\sum_{l=1}^{d^n}|c_l|^2$. We used the identity $\delta(\frac{a}{b}-1)=b\delta(a-b)$ to get the equality (\ref{A3}). Collecting factors of $r$ on the left hand side, we can use the characteristics of $\Gamma(z)=\int_{0}^{\infty}e^{-t}t^{z-1}dt,Rez>0$ to solve the problem.
	\begin{align}
	V_{\mathbb{S}^{2d^n-1}}\int_{0}^{\infty} dr r^{2d^n-1} e^{-r^2} &= \int \prod_{i=1}d^2u_i e^{-\sum_{l=1}^{d^n}|u_l|^2} \\
	V_{\mathbb{S}^{2d^n-1}}\frac{\Gamma(N)}{2} &= [\Gamma(\frac{1}{2})]^{2d^n}=\pi^{d^n} \label{A4}\\
	V_{\mathbb{S}^{2d^n-1}} &= \frac{2\pi^{d^n}}{(d^n-1)!} \label{A5}
	\end{align}
	where we have used $\Gamma(n)=(n-1)!,\Gamma(\frac{1}{2})=\sqrt{\pi}$ and $\int_{0}^{\infty}r^{2q}e^{-r^2}dr=\frac{\Gamma(q+1/2)}{2}$.
	
	Now we can calculate the correlation function for a k-body product of distinct unitary operators.
	\begin{equation}
	\begin{split}
	I(k,t) &\equiv \9 |c_1|^{2t_1}|c_2|^{2t_2}\cdots|c_k|^{2t_k} \0 \\
	&= \frac{1}{V_{\mathbb{S}^{2d^n-1}}}\int_{\mathbb{S}^{2d^n-1}}\mu_{Haar}(dU)|c_1|^{2t_1}|c_2|^{2t_2}\cdots|c_k|^{2t_k},
	\end{split}
	\end{equation}
	which corresponds to the expectation of a homogeneous monomial of degree t, where $t=\sum_{j=1}^kt_j$. Here we use $U_{|i\0}$ instead of $U_{|\psi\0}$. Similarity, we have
	\begin{equation}
	\begin{split}
	& I(k,t)V_{\mathbb{S}^{2d^n-1}}\int_{0}^\infty dr r^{2d^n-1+2\sum_{j=1}^{k}t_j}e^{-r^2}\\
	=& \int \prod_{i=1}^{d^n}d^2u_ie^{-\sum_{l=1}^{d^n}|u_l|^2}\prod_{j=1}^{k}|u_j|^{2t_j}.
	\end{split}
	\end{equation}
	Inserting the equality (\ref{A4}) and (\ref{A5}), we have
	\begin{equation}
	\begin{split}
	& I(k,t)\frac{2\pi}{(d^n-1)!}\frac{\Gamma(d^n+\sum_{j=1}^kt_j)}{2}\\
	=& [\int d^2ue^{-|u|^2}]^{d^n-k}\prod_{j=1}^k\int d^2u_je^{-|u_j|^2}|u_j|^{2t_j}.
	\end{split}
	\end{equation}
	We used $\int d^2ue^{-|u|^2}=\pi$ to get
	\begin{equation}
	I(k,t)=\frac{(d^n-1)!}{\pi^k(d^n+t=1)!}\prod_{j=1}^k\int d^2u_je^{-|u_j|^2}|u_j|^{2t_j}.
	\end{equation}
	In order to evaluate the remaining factor we change to polar coordinates, with $u_j=x+iy$, and $dxdy=rdrd\theta$, giving for each $u_j$ the factor,
	\begin{equation}
	\begin{split}
	\int d^2u_je^{-|u_j|^2}|u_j|^{2t_j} &= 2\pi\int_{0}^\infty dre^{-r^2}r^{2t_j}\\
	&= 2\pi\frac{\Gamma(t_j+1)}{2}\\
	&=\pi t_j!
	\end{split}
	\end{equation}
	Hence, 
	\begin{equation}
	\begin{split}
	I(k,t) &\equiv \9 |c_1|^{2t_1}|c_2|^{2t_2}\cdots|c_k|^{2t_k} \0 \\
	&= \frac{t_1!t_2!\cdots t_k!}{(d^n+t-1)(d^n+t-2)\cdots d^n}.
	\end{split}
	\end{equation}
	
	Following the definition of $P_{sym,t,d^n}$, we get $tr(P_{sym,t,d^n})=I(1,t)^{-1}$. Then, we have
	\begin{equation}
	\sum_{\pi\in\mathcal{S}_t}|\9\psi_{\sigma,d}|\psi_{\pi,d}\0|^n=\frac{(d^n+t-1)\cdots(d^n+1)d^n}{d^{tn}}.
	\end{equation}

\end{document}